\documentclass{article}

\usepackage{arxiv} 
\usepackage[T1]{fontenc}    
\usepackage[utf8]{inputenc} 

\usepackage[english]{babel}
\usepackage[numbers]{natbib}  

\setcitestyle{authoryear,open={(},close={)}}

\usepackage{graphicx} 
\usepackage{enumerate}
\usepackage{multirow}
\usepackage{enumitem}
\usepackage{tikz}
\usetikzlibrary{patterns}
\usepackage[most]{tcolorbox}
\usepackage{listings}

\usepackage{hyperref}       
\usepackage{url}            
\usepackage{hyperref}
\hypersetup{
	colorlinks=true,
	linkcolor=blue,
	filecolor=blue,
	citecolor = blue,      
	urlcolor=cyan,
}
\usepackage{booktabs}       
\usepackage{multicol}
\usepackage{amsmath,amsfonts,amssymb}
\usepackage{comment}
\usepackage[normalem]{ulem}

\usepackage{amsthm}         
\numberwithin{equation}{section}

\usepackage{nicefrac}       
\usepackage{microtype}      
\usepackage{lipsum}
\usepackage{float} 
\usepackage{bm}
\usepackage{adjustbox}
\usepackage[linesnumbered,lined,boxed,commentsnumbered]{algorithm2e}
\usepackage{soul}
\usepackage{graphicx}
\usepackage{subcaption}

\newcommand{\twee}{\operatorname{Tweedie}}

\newcommand{\Esp}[1]{\mathrm{E}\! \left[ #1 \right]}
\newcommand{\Var}[1]{\mathrm{Var}\! \left[ #1 \right]}

\newcommand{\Prr}[1]{\mathrm{Pr}\! \left( #1 \right)}

\usepackage{caption}

\usepackage{tikz}
\usetikzlibrary{shapes,arrows}

\usetikzlibrary{decorations.pathreplacing}

\newtcolorbox{hatchedblock}{
  enhanced,
  sharp corners,
  colback=white,     
  boxrule=0pt,       
  frame hidden,
  interior engine=empty,
  overlay={
    \begin{scope}
      \clip (interior.south west) rectangle (interior.north east);
      \path[pattern=north east lines,pattern color=black]
        (interior.south west) rectangle (interior.north east);
    \end{scope}
  }
}

\title{Varying Risk Exposure in Auto Insurance: A Weighted Tweedie Framework for Experience Rating and Cancellation Penalties}

\author{
Jean-Philippe Boucher$^{1}$,
Raïssa Coulibaly$^{1}$,
Julien Trufin$^{2}$ \\[0.3cm]
$^{1}$ Chaire Co-operators en analyse des risques actuariels,
Département de mathématiques, UQAM, Montréal, Canada \\
$^{2}$ Department of Mathematics, Université Libre de Bruxelles (ULB), Bruxelles, Belgium
}

\begin{document}
\maketitle

\begin{abstract}
This paper proposes a new family of Tweedie-based ratemaking models that explicitly account for mid-term policy cancellations. Using an automobile insurance dataset from a Canadian insurer, we document a marked difference in claims experience between policyholders who maintain their coverage until maturity and those who cancel their policies mid-term. Building on the classical Tweedie framework, we introduce flexible weighting functions and a premium penalty structure that depend on the level of exposure, allowing for a more realistic representation of the earned premium when coverage is interrupted before the end of the policy period. We compare several weighting structures within the Tweedie framework and examine their theoretical properties, as well as their empirical performance using deviance-based model comparison criteria, an area-between-curves criterion derived from concentration and Lorenz curves, and Murphy diagrams grounded in Bregman dominance. To operationalize the proposed models, monotonicity and non-negativity constraints are imposed on the penalty function, ensuring consistency with actuarial principles. Finally, using real-world data, we show that this approach provides both a strategic and competitive advantage: it allows the insurer to indirectly compensate for large losses through a cancellation surcharge, while preserving actuarial coherence and statistical consistency.
\end{abstract}


\newpage 
\section{Introduction}\label{introduction}

In property and casualty insurance—automobile insurance in particular—contracts are typically written for a one-year term. At renewal, a policyholder may either remain with the same insurer or move to a competitor. Although most policy modifications occur at renewal, mid-term cancellations—situations in which a policy is terminated before its scheduled expiry date—are also common in North America. Insurers often levy a financial penalty in such cases, primarily to recover administrative costs. Several situations may lead to mid-term cancellations:

\begin{enumerate}
    \item The insured sells the vehicle, and coverage is no longer required;
    \item The insured experiences a major loss (e.g., theft or total loss) and replaces the vehicle, thereby requiring a new policy;
    \item The insured switches to another insurer, often motivated by lower premiums despite the cancellation charge.
\end{enumerate}

Cases (1) and (2) are sometimes treated operationally as vehicle substitutions rather than true cancellations, and insurers frequently waive cancellation penalties when coverage continues for the replacement vehicle. From an actuarial standpoint, however, substitutions introduce important data and modelling challenges. Standard pricing databases are not designed to track vehicle swaps: the cancellation of one contract and the creation of another often occur at different dates (sometimes weeks apart), and the designation of the principal driver may change across vehicles within the same household. These timing inconsistencies complicate the identification of substitutions and hinder the construction of coherent longitudinal exposure histories.

Because commonly used pricing datasets do not explicitly record substitutions, actuaries typically treat each vehicle contract independently and rely solely on the contract’s recorded exposure duration. Recent work has begun to exploit dependencies among vehicles held by the same policyholder \citep{raissaboucher2024, turcotte2022gamlss, pechon2019multivariate}, but these approaches do not fully resolve the practical difficulties created by asynchronous cancellations and additions. Until richer, transaction-level data become widely available, pricing practice will continue to rely on modelling each contract’s observed exposure.

It is therefore essential to distinguish contracts cancelled mid-term from those affected by vehicle additions or replacements. Although both contract types have an exposure duration shorter than one year, they arise from different operational mechanisms. As detailed in Section~\ref{SectData}, mid-term cancellations correspond to contracts whose termination date differs from the initially scheduled end date, whereas vehicle additions and replacements retain the original policy end date. Regardless of the underlying cause, accurate modelling of such contracts requires careful consideration of risk exposure.

However, most pricing models that account for exposure rely on the assumption that the premium $\mu$—defined as the expected value of the response variable—is proportional to the exposure $t$, typically specified as (see, e.g., \cite{denuit2007actuarial, frees2014predictive})

\begin{equation}\label{meanintro}
    \mu = t\,\exp\left\{\mathbf{x}^{\top}\boldsymbol{\beta}\right\},
\end{equation}

where \(\boldsymbol{\beta}\) denotes the vector of parameters to be estimated and $\mathbf{x}$ the covariate vector. In actuarial science, loss costs are commonly modeled using distributions from the Tweedie family (see \cite{tweedie1984index}), which is parameterized by two components through which exposure may affect premiums: the mean $\mu$ and the weight parameter $w$. In this context, \cite{raissaboucher2025} examine two classical specifications—the offset formulation ($w = 1$) and the ratio formulation ($w = t^{p-1}$)—both grounded in the proportional mean structure given in \eqref{meanintro}. Their analysis establishes that the offset approach is asymptotically more efficient than the ratio approach. However, when performance is assessed in terms of portfolio-level financial equilibrium, the ratio approach proves to be more effective. Although these findings are insightful, the underlying proportionality assumption between premiums and exposure appears overly restrictive. Unlike \cite{raissaboucher2025}, who retain the proportional mean structure and modify only the weight specification, we relax the proportionality assumption itself and allow the exposure to enter the mean through a flexible function. In particular, the empirical evidence reported in Section~\ref{SectData} challenges this assumption. The objective of the present study is therefore twofold:

\begin{enumerate}
    \item To generalize the specification of the mean premium \(\mu\) to allow more flexible relationships between exposure $t$ and expected loss, i.e. $\mu = \gamma(t)\,\exp\left\{\mathbf{x}^{\top}\boldsymbol{\beta}\right\}$;
    \item To identify an appropriate weight function \(w\) consistent with this generalized structure. 
\end{enumerate}

This extension is motivated by both empirical evidence and practical considerations in premium calculation and risk classification.  It is also worth noting that mid-term cancellations resulting from major losses (such as total losses requiring vehicle replacement) may create additional strategic considerations for insurers. In such cases, insurers may benefit from structuring cancellation penalties to partially reflect the occurrence of severe claims. This mechanism effectively allows the insurer to indirectly identify major losses through the cancellation surcharge, capturing part of the expected cost associated with the claim. From a competitive perspective, this ensures that policyholders maintaining full-year coverage implicitly benefit from a reduced annual premium, whereas those cancelling mid-term bear a larger share of their corresponding expected cost. Consequently, it is important to consider what adjustments may be introduced to accommodate this flexible framework.

The structure of the paper is as follows. Section~\ref{SectData} presents a detailed analysis of an automobile insurance dataset from a Canadian insurer and highlights the strong relationship between mid-term cancellations and poor driving behavior, thereby motivating the need for a more flexible pricing framework. Section~\ref{Sect3} introduces several practical strategies designed to better penalize mid-term cancellations. Section~\ref{Tweediedistribution} formally develops the Tweedie distribution theory underlying our ratemaking models. Section~\ref{sectNum} illustrates the proposed methods using a numerical case study based on the same dataset, assessing their impact on insurer profitability and risk selection. Section~\ref{Conclu} concludes.

\section{Data Description and Objectives} \label{SectData}

We employ a dataset constructed from a non-random sample of an automobile insurance portfolio provided by a major Canadian insurer. The data cover a period of 13 consecutive years and pertain to the province of Ontario. In total, the dataset contains more than two million records, each corresponding to an annual insurance contract for a single vehicle. For every contract, the database includes key identifiers such as the policy number, vehicle identification, and contract inception and termination dates.

Although the dataset spans 13 consecutive years, the first six years are used to construct the policyholders’ claims history, whereas the remaining seven years constitute the analysis sample, which is subsequently randomly partitioned for estimation, model selection, and out-of-sample validation into a training set ($75\%$) and a test set ($25\%$).

In addition to contractual details, the dataset contains rich information on policyholders and their vehicles, as well as the frequency and severity of claims associated with each contract. Claims are further classified by coverage type, including third-party liability, collision, and comprehensive insurance. In this study, we restrict attention to collision coverage, which insures property damage to the policyholder’s vehicle in at-fault accidents.

In order to inform the specification of the mean function $\mu$ introduced in the previous section, we begin by describing the dataset and by presenting exploratory analyses that highlight the empirical relationship between recorded exposure and aggregate cost. The primary aim of this section is to document the data and the main variables used in the study, and to provide preliminary evidence on the functional form of $\mu$ that will guide subsequent modeling choices (e.g. offset vs. ratio vs. flexible exposure functions, and the selection of an appropriate weighting scheme).

\subsection{Description of Contracts}

To comprehensively account for all situations involving mid-term cancellations, we classify all contracts in the dataset into four distinct groups according to their exposure period:  

\begin{itemize}
    \item $\mathcal{XX}$: Contracts with full risk exposure for the entire duration of the policy.
    \item $\mathcal{OX}$: Contracts initiated during a policy period and exposed to risk until the policy expires.
    \item $\mathcal{XO}$: Contracts exposed to risk at the start of a policy period but canceled before the policy expires.
    \item $\mathcal{OO}$: Contracts initiated during a policy period and canceled before the policy expires.
\end{itemize}

\begin{figure}[h]
\begin{adjustbox}{max width=\textwidth}
\begin{tikzpicture}[scale=0.85, auto, to/.style={->,>=stealth',shorten >=1pt}, every node/.style={font=\fontsize{9pt}{9pt}\selectfont\sffamily, align=center, semithick}]
\draw[-] (0,0) -- (17,0);
\draw[-] (0,4) -- (17,4);

\draw[dashed] (0, 0) -- (0, 4);
\draw[dashed] (17, 0) -- (17, 4);

\draw (-0.75,3.2) node[above=0pt] {$\mathcal{X}\mathcal{X}$};
\draw[-] (0,3.5) -- (17,3.5);
\filldraw[black,fill=black] (0,3.5) circle (3pt);
\filldraw[black,fill=black] (17,3.5) circle (3pt);

\draw (-0.75,2.2) node[above=0pt] {$\mathcal{O}\mathcal{X}$};
\draw[-] (3,2.5) -- (17,2.5);
\filldraw[red,fill=red] (3,2.5) circle (3pt);
\filldraw[black,fill=black] (17,2.5) circle (3pt);

\draw (-0.75,1.2) node[above=0pt] {$\mathcal{X}\mathcal{O}$};
\draw[-] (0,1.5) -- (13,1.5);
\filldraw[black,fill=black] (0,1.5) circle (3pt);
\filldraw[red,fill=red] (13,1.5) circle (3pt);

\draw (-0.75,0.2) node[above=0pt] {$\mathcal{O}\mathcal{O}$};
\draw[-] (3,0.5) -- (13,0.5);
\filldraw[red,fill=red] (3,0.5) circle (3pt);
\filldraw[red,fill=red] (13,0.5) circle (3pt);

\foreach \x in {0, 17}   \draw (\x cm,3pt) -- (\x cm,-3pt);

\draw (0,-0.5) node[below=1pt] {Start of the Policy Period};
\draw (17,-0.5) node[below=1pt] {End of the Policy Period};

\end{tikzpicture}
\end{adjustbox}
\caption{\textit{All truncation possibilities for a one-year policy period}}
\label{CasesOneYear}
\end{figure}
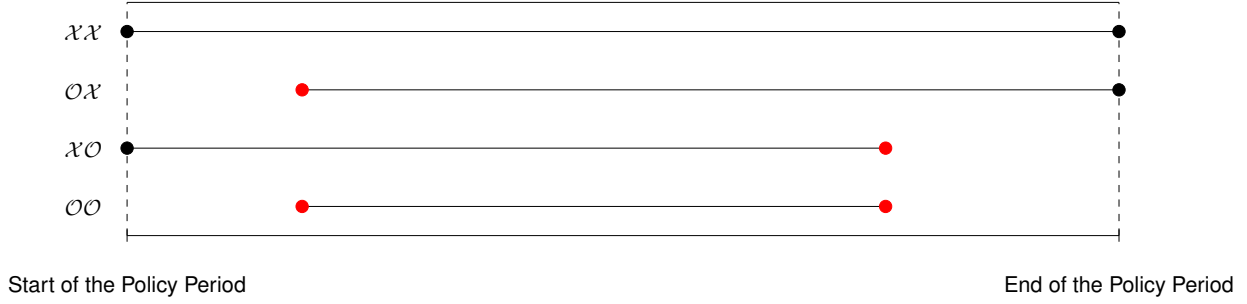

These groups can be better understood by referring to Figure \ref{CasesOneYear}. This classification provides a structured framework to capture the effects of mid-term cancellations and vehicle substitutions, thereby facilitating a more accurate analysis of insurance contract dynamics. However, since our objective is to model the penalty applied to policyholders who cancel before the end of their contract—rather than to address cases where additional vehicles (or coverages) are added during the term—we restrict our analysis to policyholders of types $\mathcal{XX}$ and $\mathcal{XO}$ only. Table \ref{Anum.tabl333} reports the proportions of policyholders in the $\mathcal{XX}$ and $\mathcal{XO}$ classes for both the training and testing datasets.

\begin{table}[H]
    \centering
    \begin{tabular}{l|cc}
\hline
Datasets & \multicolumn{2}{c}{Contract type} \\
         & $\mathcal{XX}$ & $\mathcal{XO}$ \\
\hline
Training set & 323\,231 (34.7\%) & 608\,261 (65.3\%) \\
Test set     & 138\,852 (34.8\%) & 260\,360 (65.2\%) \\
\hline
\end{tabular}
    \caption{Descriptive statistics by contract type}
    \label{Anum.tabl333}
\end{table}

\subsubsection{Distribution of Exposure}

\begin{figure}[H]
	\begin{center}
		\includegraphics[scale=0.54]{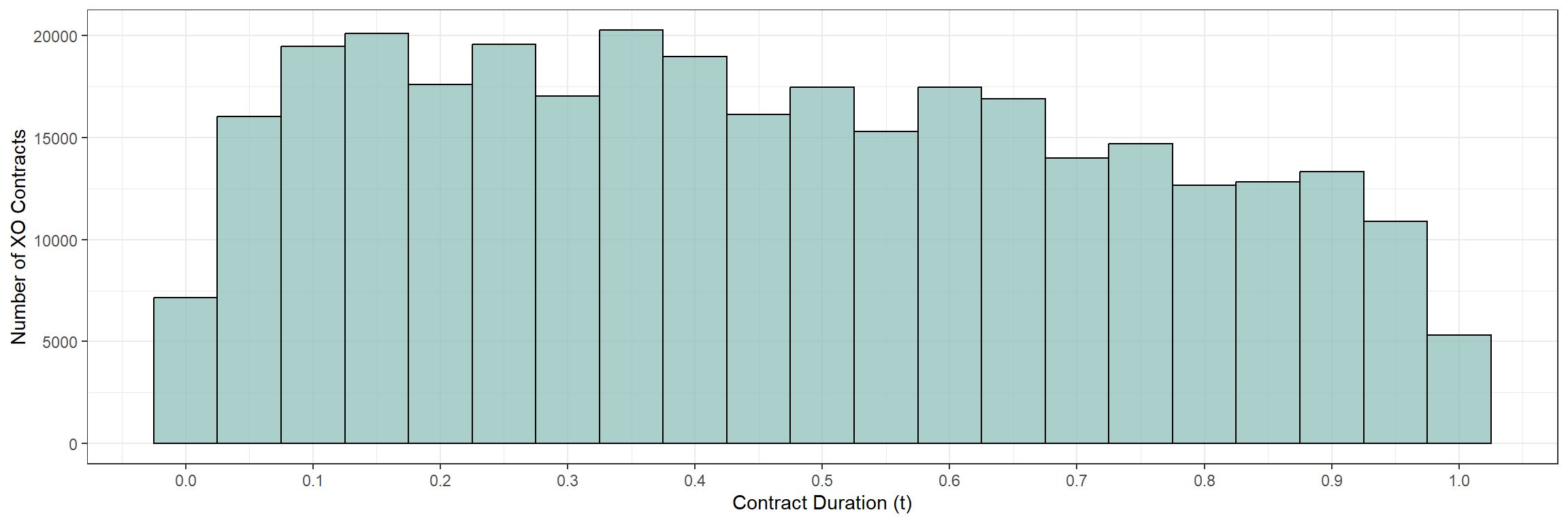}
	\caption{Distribution of the risk exposure for the $\mathcal{XO}$ group}
		\label{histoexposure}
	\end{center}
\end{figure}

The histogram of Figure \ref{histoexposure} presents the distribution of risk exposure specifically for the $\mathcal{XO}$ group in the training set, i.e., contracts that were canceled before the end of the policy period. All full-year exposures are excluded since they correspond to the $\mathcal{XX}$ group. This visualization highlights the variation in partial-year exposures and provides insight into the timing and prevalence of mid-term cancellations.

\subsection{Empirical Impact of Mid-Term Cancellations on Contract Risk} \label{empiricalanalysis}

To better understand the impact of mid-term cancellations on contract risk, and to challenge the standard pricing assumption of linearity between contract duration (risk exposure) and the premium, we first recall the definition of the loss cost, defined as the total amount paid in claims over the coverage period.

Consider $n$ insurance contracts. Let $n_i$ denote the total number of claims reported for contract $i$, for $i=1,\ldots,n$, and let $z_{i,k}$ denote the cost (i.e., claim severity) of the $k$-th claim, for $k=1,\ldots,n_i$. The observed loss cost for contract $i$, denoted by $y_i$, is therefore given by
\[
y_i=\begin{cases}
\sum_{k=1}^{n_i} z_{i,k}, & \text{if } n_i>0,\\[0.2cm]
0, & \text{if } n_i=0.
\end{cases}
\]

Accordingly, a first way to quantify the risk associated with a contract using the total loss cost is to compute the average loss cost, defined as $\frac{1}{n}\sum_{i=1}^n y_i$. However, since some contracts may terminate before the end of the coverage period, it is more appropriate to consider an exposure-adjusted indicator, referred to as the average annualized loss cost, defined as $\frac{1}{\sum_{i=1}^n t_i}\sum_{i=1}^n y_i$, where $t_i$ denotes the risk exposure of contract $i$. Using the definition of the loss cost, the average annualized loss cost can be expressed as
\[
\frac{1}{\sum_{i=1}^n t_i}\sum_{i=1}^n y_i
= \left( \frac{\sum_{i=1}^n n_i}{\sum_{i=1}^n t_i} \right)
\left( \frac{\sum_{i=1}^n \sum_{k=1}^{n_i} z_{i,k}}{\sum_{i=1}^n n_i} \right),
\]

where, by convention, $\sum_{k=1}^{n_i} z_{i,k} = 0$ when $n_i = 0$, corresponding to the product of the claim frequency and the average claim severity.

\begin{figure}[H]
	\begin{center}
		\includegraphics[scale=0.54]{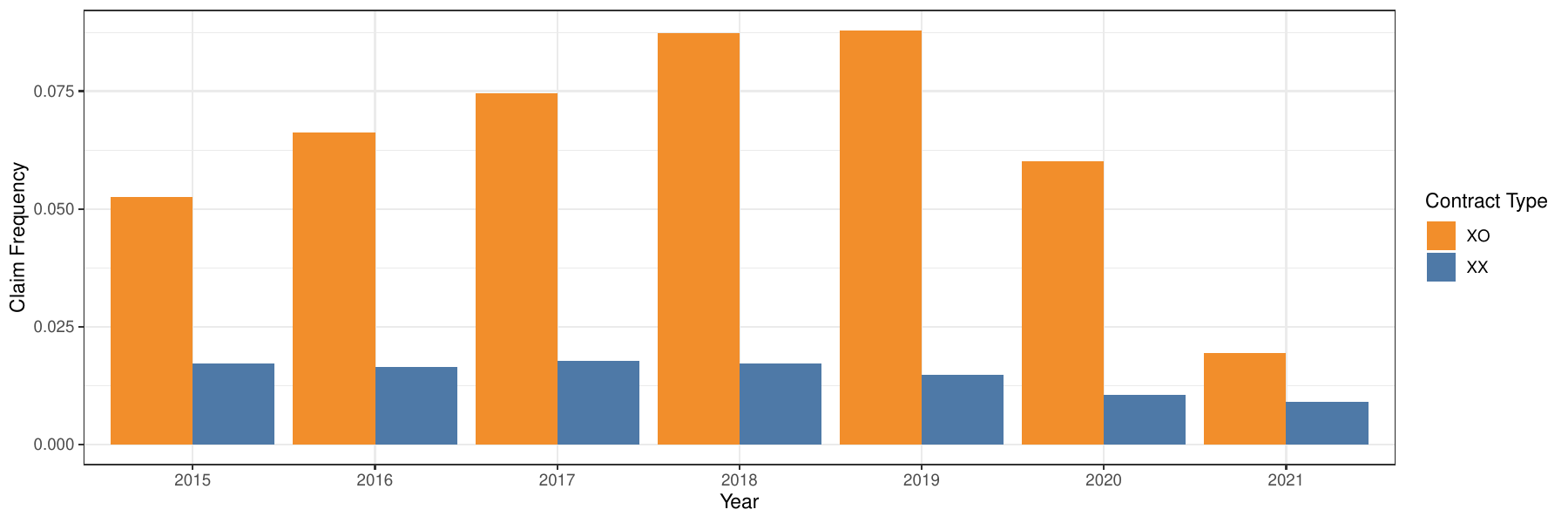}
	\caption{Claim frequency by policy year and contract type}
		\label{Freqbyexposure}
	\end{center}
\end{figure}

Our empirical strategy therefore consists in analyzing each component of this decomposition in order to better characterize contracts that are canceled mid-term. It is also important to note that this analysis is conducted using the training sample. We begin by examining the claim frequency, defined as the ratio of the total number of claims to the total exposure, namely $\frac{\sum_{i=1}^n n_i}{\sum_{i=1}^n t_i}$. More specifically, we study how this indicator varies by policy year and by contract type. Figure~\ref{Freqbyexposure} presents the corresponding empirical distribution based on the training sample from the seven-year analysis period (i.e., the last seven years of the dataset). Although a more detailed model will eventually need to incorporate policyholder characteristics, we can already observe, consistently across years, that the claim frequency of policyholders who cancel their contracts mid-term is much higher than that of policyholders who maintain coverage until the end of the policy period.

A similar analysis can be conducted for the average claim severity component. However, since claim severity is a continuous variable, it is particularly informative to examine its distribution across the two contract groups. This distributional perspective provides additional insights beyond those obtained from the claim frequency analysis.

\begin{figure}[H]
	\begin{center}
		\includegraphics[scale=0.54]{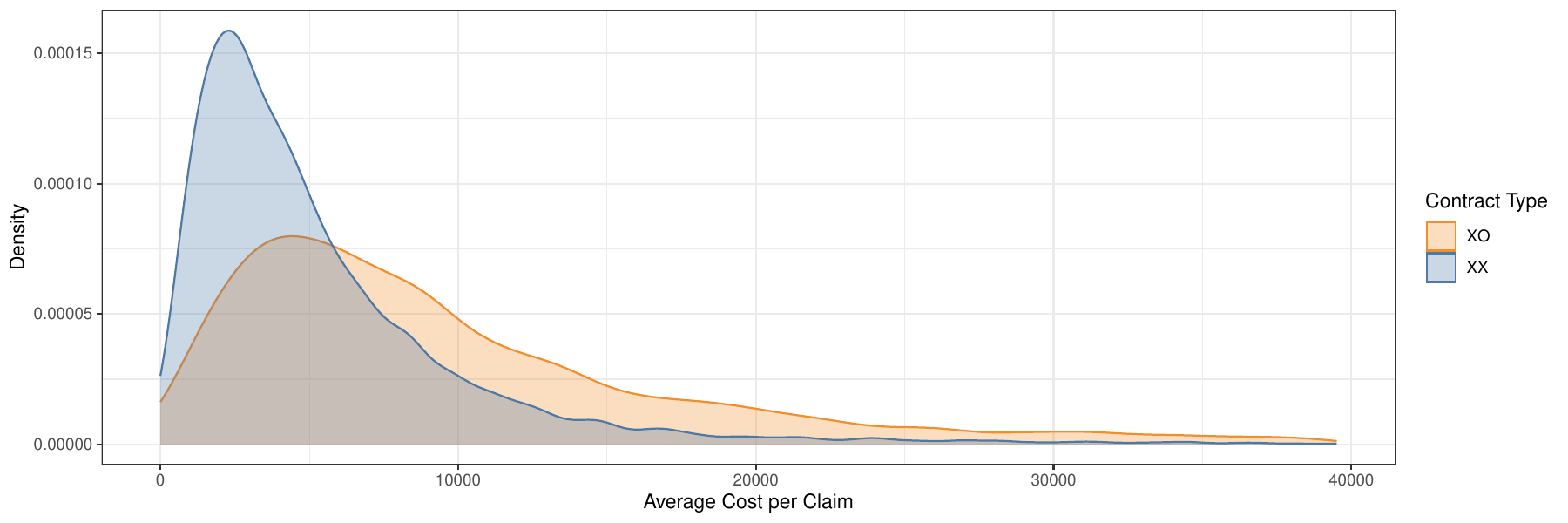}
	\caption{Average cost per claim by contract type}
		\label{Sevbyexposure}
	\end{center}
\end{figure}

Accordingly, for each group, we compute the average cost per claim, defined as $\frac{1}{n_i}\sum_{k=1}^{n_i} z_{i,k}$ when $n_i>0$, and estimate the corresponding kernel density, as illustrated in Figure~\ref{Sevbyexposure}. Interestingly, policyholders who cancel mid-term not only exhibit a higher claim frequency (relative to their risk exposure) but also tend to have higher claim severity, which further suggests that more severe claims may trigger policy cancellations.

Finally, the two figures below display the average loss cost and the average annualized loss cost as functions of risk exposure (with all policyholders in the $\mathcal{XX}$ group grouped at an exposure level of 1). For these plots, exposure is discretized into intervals of width 0.05, and policyholders are aggregated within each interval. The size of each point reflects the number of policyholders in the corresponding exposure group. Consistent with the patterns observed in the analyses of claim frequency and cost per claim, contract duration has a clear impact on claim outcomes.

\begin{figure}[H]
	\begin{center}
		\includegraphics[scale=0.54]{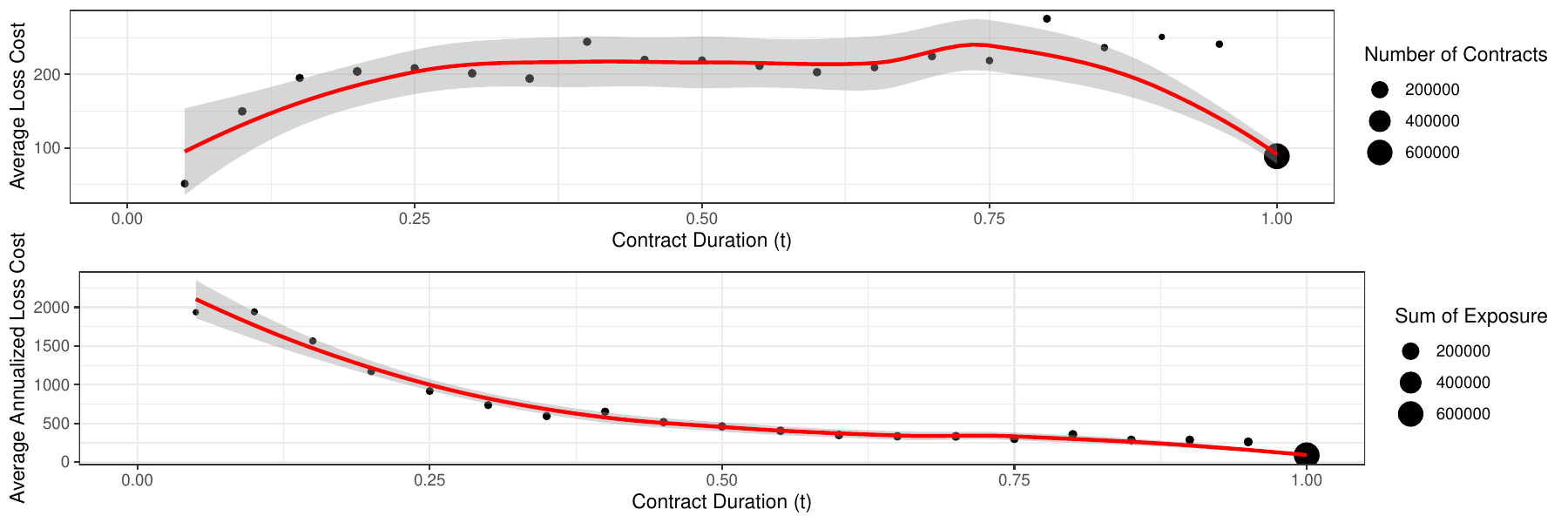}
	\caption{Average loss cost by risk exposure}
		\label{LCbyexposure}
	\end{center}
\end{figure}

\subsection{Available Covariates}

The dataset includes multiple attributes for each contract and vehicle. To investigate how segmentation affects pricing, we focus on five primary covariates, labeled $X_1$ through $X_5$ for confidentiality reasons. These variables represent common risk factors, such as policyholder characteristics, vehicle type, and usage patterns, in line with conventional actuarial practice. For the sake of model parsimony, we exclude features that are not typically incorporated into standard pricing frameworks, so that the resulting specification remains consistent with widely used industry methodologies. Summary statistics for these covariates are provided in Appendix I.

It should be emphasized, however, that the selected covariates are not intended to be fully representative of those commonly used in practice, as their detailed impact is not the primary focus of our analysis. In particular, our objective is not to assess finely the contribution of each covariate to risk differentiation, but rather to use them as illustrative controls within the modeling framework.

Rather than focusing solely on understanding risk exposure and mid-term cancellations in the portfolio as a function of standard covariates, we place greater emphasis on the individual policyholder’s experience with the insurer and their claims history. While it is clear that conventional covariates, such as age and gender, vehicle type, and usage patterns, affect the modeling of risk exposure, we argue that the policyholder’s behavioral and historical profile is even more informative.

\subsection{Claims Experience}\label{SectDataBMS}

Policyholders with higher inherent risk tend to report claims more frequently, which increases not only their expected loss cost but also the likelihood of experiencing a severe claim that may lead to a mid-term contract termination. To better distinguish among policyholders with different risk profiles, we follow the approach proposed by \cite{raissaboucher2024} and focus on a key covariate available in the dataset: the Bonus--Malus Scale (BMS) level, denoted by $\ell$. By summarizing past claims experience, this level provides a structured and interpretable measure of individual risk, which is essential for isolating the behavior of exposure groups such as $\mathcal{XX}$ and $\mathcal{XO}$.

In practice, insurers routinely adjust premiums based on past claims, rewarding claim-free histories and penalizing frequent claimants. The BMS formalizes this mechanism by assigning each policyholder a level $\ell$ according to their historical claims. This level is updated at each contract renewal, increasing after a claim and decreasing following a claim-free period. Consequently, the BMS condenses a policyholder’s entire claims record into a single, actuarially meaningful indicator of risk.

\begin{figure}[H]
	\begin{center}
		\includegraphics[scale=0.54]{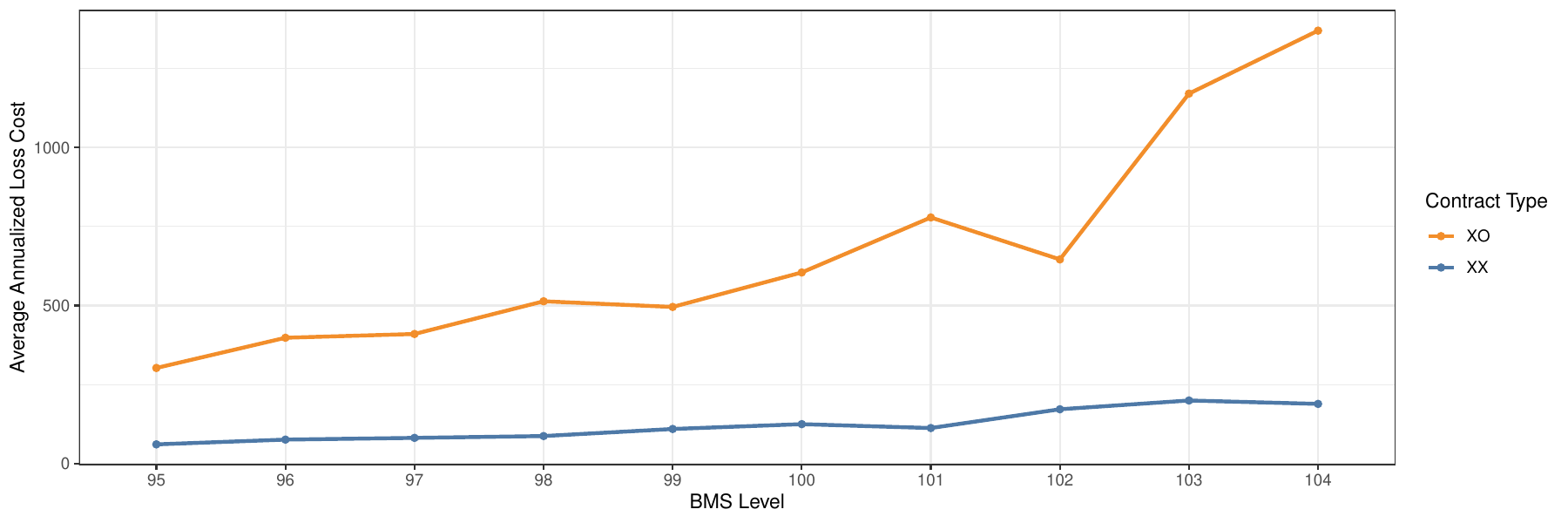}
	\caption{Average annualized loss cost by BMS level and contract type}
		\label{BMSLS}
	\end{center}
\end{figure}

In the context of analyzing mid-term cancellations, the BMS level is particularly informative: it not only reflects underlying claim risk but may also influence the probability of early contract termination. Because the BMS parameters for this portfolio were already calibrated and optimized in \cite{raissaboucher2024}, we directly use the resulting BMS levels as observed covariates. The system applied in the dataset is characterized by an initial level of 100 for new policyholders, a jump parameter of 3 for each reported claim, and admissible levels ranging from 95 to 104. Thus, policyholders with few or no claims tend to accumulate levels close to $\ell = 95$, whereas those with multiple claims can reach levels up to $\ell = 104$; policyholders without significant claims generally remain around $\ell = 100$. This relationship is illustrated in Figure~\ref{BMSLS}, which displays the average annualized loss cost across BMS levels separately for contracts of type $\mathcal{XX}$ and $\mathcal{XO}$.

Consistent with the findings of \cite{raissaboucher2024}, the figure reveals a clear positive association between the BMS level and the average annualized loss cost: lower BMS scores are associated with lower annualized loss costs, whereas higher BMS scores correspond to higher annualized loss costs. In addition, a pronounced separation between the two contract types is observed. Across all BMS levels, contracts of type $\mathcal{XX}$ exhibit consistently lower risk levels than contracts of type $\mathcal{XO}$, which are characterized by higher annualized loss costs and a greater propensity for mid-term cancellation.

\begin{figure}[H]
	\begin{center}
		\includegraphics[scale=0.54]{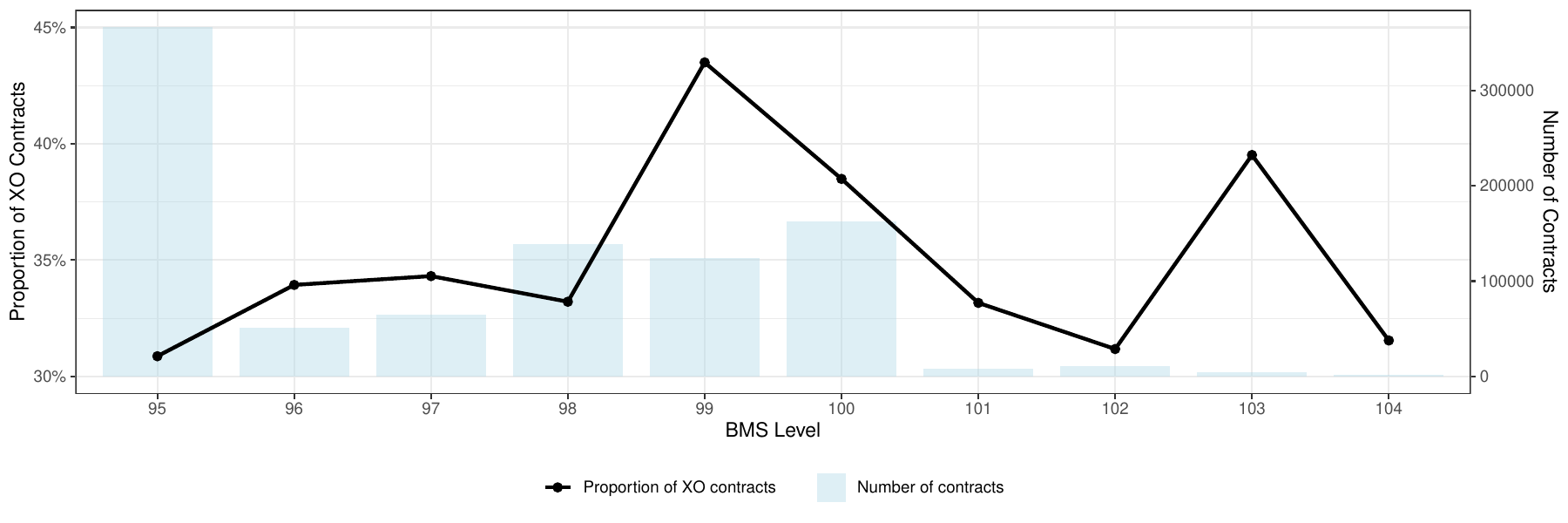}
	\caption{Proportion of $\mathcal{XO}$ contracts by BMS level and number of contracts}
		\label{BMSProp}
	\end{center}
\end{figure}

To further characterize the relationship between the BMS score and contract type, Figure~\ref{BMSProp} displays the proportion of $\mathcal{XO}$ contracts across BMS levels, represented by the black line, while the light-blue bars indicate the corresponding number of contracts. A large share of policyholders is concentrated at $\ell = 95$, the most favorable level, reflecting a substantial segment of low-risk individuals. This group exhibits the lowest proportion of $\mathcal{XO}$ contracts. At the opposite end of the spectrum, policyholders with levels $\ell > 100$ constitute a relatively small portion of the portfolio and are generally considered the riskiest. Interestingly, their proportion of mid-term cancellations is not particularly high. Conversely, policyholders at intermediate levels, particularly $\ell = 100$ (often corresponding to newer clients) and $\ell = 99$ (those with limited claims history) display the highest proportion of $\mathcal{XO}$ contracts. Although the distribution of $\mathcal{XO}$ contracts across BMS levels appears relatively homogeneous, it remains informative regarding the likelihood of future contract cancellation. In particular, policyholders who are aware that a recent claim may substantially increase their future premium — due to the penalty induced by a deterioration in their BMS level — may be more inclined to terminate their contract following such an event. Moreover, the BMS score may implicitly capture additional behavioral factors influencing the decision to cancel coverage. It is also important to note that, while the BMS level is treated as an exogenous covariate in our empirical analysis, it is inherently endogenous to the underlying pricing and claims process.

\section{Characterizing the Mean Response to Risk Exposure}
\label{Sect3}

\subsection{Ratemaking Strategies} 

When incorporating the risk exposure into premium calculations, two main approaches may be considered:

\begin{enumerate}
    \item \textbf{Traditional Approach}: For contracts that terminate before the end of the policy year, the conventional actuarial assumption is that the expected claim amount is proportional to exposure. Denoting by $t$ the fraction of the year during which the contract was active, the standard specification is:
    \[
    \mu = t \times \exp\left\{\mathbf{x}^{\top}\boldsymbol{\beta}\right\} \equiv \pi^{\mathrm{Trad}}(t),
    \]
    where $\exp \big\{\mathbf{x}^\top \bm{\beta}\big\}$ represents the annualized premium corresponding to a full year of coverage for a policy with covariate vector $\mathbf{x}$. This formulation implies that, in the absence of administrative fees, an insured who cancels halfway through the year would receive a pro rata refund equal to one half of the annual premium—that is, the premium scales linearly with recorded exposure $t$.

    \item \textbf{Flexible Approach}: Empirical observations from the data suggest that the relationship between risk exposure $t$ and the expected loss cost may be more complex than assumed under the traditional approach. Motivated by this finding, we consider the following more flexible specification:
    \[
    \mu = \gamma(t) \exp\left\{\mathbf{x}^{\top}\boldsymbol{\beta}\right\} \equiv \pi^{\text{Flex}}(t),
    \]
    where $\gamma(t)$ is a smooth exposure function capturing deviations from linearity. Estimating $\gamma(t)$ directly can be challenging. As demonstrated in the Numerical Application of Section~\ref{sectNum}, we propose to use the framework of Generalized Additive Models (GAMs), following the approach introduced by \cite{wood2017}. GAMs extend generalized linear models by incorporating non-linear relationships through smooth functions, making them well suited for capturing complex exposure–risk patterns commonly observed in insurance data.

    Adopting a flexible exposure function is particularly appealing for insurers, as it enables a more refined treatment of mid-term cancellations—especially those arising after major loss events. By allowing differentiated refunds or surcharges that more accurately reflect actual exposure, such a specification improves the alignment between premium and risk, reduces cross-subsidization among policyholders, and may ultimately confer a competitive advantage.
\end{enumerate}

\subsection{Penalty Structures}\label{sectPenalty}

While the flexible approach extends the traditional one and is likely to provide a better fit to the data—both on the training set and on new data, as suggested by our numerical application—this does not necessarily imply that it represents the optimal choice in practice. The actual risk exposure of a vehicle under a given contract is unknown at the time the policy is issued, and regulatory authorities often require insurers to disclose the full premium upfront, even if the contract is later canceled mid-term. Such requirements aim to prevent non-transparent pricing practices, for example those that could disproportionately penalize policyholders who cancel after experiencing an accident. However, these regulations do not completely rule out the possibility of implementing more sophisticated penalty structures. If all potential penalties are clearly stated in the contract prior to subscription, the insurer could, at least in principle, apply a premium scheme that better reflects expected risk exposure, even when cancellation occurs early. Within this framework, contractual transparency requirements may still allow insurers to design more adequate and actuarially grounded penalty structures, which is the perspective adopted in the analysis below.

For simplicity and to streamline notation, let us first assume that the policyholder pays the full annual premium $\pi^{\text{Flex}}(1)$, denoted by $\pi^{\text{Flex}}$, at the start of the contract. Under standard industry conventions, if the policyholder cancels the contract at time $t$, they expect a refund equal to $(1-t)\pi^{\text{Flex}}$. However, in a flexible pricing framework where risk exposure $t$ is incorporated into the premium through a function $\gamma(t)$, a cancellation occurring at time $t$ would typically imply a refund of $ \pi^{\text{Flex}} - \gamma(t) \pi^{\text{Flex}}$. This model-based refund does not necessarily coincide with the conventional refund.

To evaluate whether it is at least partially feasible to implement a premium structure that explicitly accounts for mid-term cancellations, we compare these two refund amounts: the refund expected by the policyholder and the refund generated by the flexible model. We define the difference between these quantities as the \emph{cancellation penalty}, denoted by $\rho(t)$:

\begin{align*}
\rho(t) &= \gamma(t) \pi^{\text{Flex}} - t \pi^{\text{Flex}}  \\
&= \pi^{\text{Flex}} (\gamma(t)  - t). 
\end{align*}

Intuitively, the premium $\pi^{\text{Flex}}(t)$ and the penalty $\rho(t)$ associated with a cancellation at time $t$ should satisfy some fundamental principles. To formalize these requirements, we impose the following two constraints:

\begin{itemize}
\item \textbf{(C1)}: $\pi^{\text{Flex}}(t) \ge \pi^{\text{Flex}}(t')$ for all $t \ge t'$, meaning that the refund prescribed by the flexible approach decreases as the contract progresses. Equivalently, the function $\gamma(t)$ must be increasing in the risk exposure $t$.

\item \textbf{(C2)}: $\rho(t) \ge 0$, ensuring that the cancellation penalty is non-negative. At a minimum—and for strategic consistency—the premium charged for coverage lasting until time $t$ should match the conventional benchmark $t \pi^{\text{Flex}}$ that a policyholder would naturally expect. Furthermore, the pricing model must be protected against offering mid-term premiums below this benchmark, as doing so would undermine the intended structure and artificially inflate the annual premium.
\end{itemize}

One may further argue that the penalty never exceeds the remaining portion of the premium. This condition guarantees that a policyholder never pays more by canceling than by maintaining the contract until its maturity. However, given that  the function $\gamma(t)$ must be increasing in the risk exposure $t$ in (C1), this requirement is automatically satisfied. 

As we will show in the empirical application, analyzing the premium $\pi^{\text{Flex}}(t)$ and the penalty $\rho(t)$ remains essential, and in some situations it may even be desirable to transform the penalty structure so as to ensure logically coherent and actuarially meaningful values. This point is crucial: imposing penalty structures different from the conventional one may alter policyholder behavior. By analogy with Lemaire’s \emph{bonus-hunger effect} \citep{lemaire}, which highlights how changes in claim-related penalties affect subsequent claim frequency, one must recognize that modifying penalties for mid-term cancellations will influence both the realized claims experience as a function of $t$  and the cancellation probability at duration $t$. 

The central idea is that insurers could, in principle, adopt alternative mid-term cancellation penalty structures. More generally, suppose that the current penalty function $\rho(t)$ is replaced by a modified structure $\rho'(t)$. Under such a change, the premium charged to a policyholder for coverage lasting a duration $t$ could take the form
\[
t\,\pi^{\mathrm{Flex}} + \rho'(t),
\]
which naturally raises the question of how to quantify the modified penalty $\rho'(t)$. This specification will be examined in detail in Section~\ref{sectNum}.

\section{Understanding the Tweedie Distribution in Actuarial Models} \label{Tweediedistribution}

To operationalize the ratemaking strategies and penalty structures discussed in Section~\ref{Sect3}, it is necessary to specify an appropriate statistical distribution for the response variable in the proposed models. As motivated in the Introduction, we adopt the Tweedie distribution for this purpose. The objective of this section is therefore to provide theoretical justification for the performance of the flexible approach under the assumption that the loss cost follows a Tweedie distribution. Before presenting these results, we first discuss the main reasons for the widespread use of the Tweedie distribution in insurance pricing, in order to build intuition for the subsequent theoretical developments.

\subsection{Justification in Auto Insurance Pricing}

\subsubsection{Variance–Mean Relationship}

The Tweedie distribution is commonly characterized by four parameters: the mean parameter, denoted by $\mu$, which represents the expected value of the distribution; the dispersion parameter, denoted by $\phi$; the weight parameter, denoted by $w$; and the variance parameter, denoted by $p$. While the dispersion and weight parameters must be strictly positive, the mean parameter is, in principle, real-valued. However, in insurance pricing applications, the mean parameter is required to be positive, and additional restrictions are typically imposed on the variance parameter $p$, as discussed in the next subsection.

A key feature of the Tweedie distribution in insurance pricing is the functional relationship linking its variance to its mean parameter. More formally, if a random variable $Y$ follows a Tweedie distribution, its variance is given by
\[
\Var{Y} =  \frac{\phi}{w}\mu^{p}.
\]

When the response variable in a pricing model follows a Tweedie distribution, this variance specification implies a positive relationship between the mean parameter $\mu$ (interpreted as the premium) and the variability of the response. This feature allows the model to reflect the empirically observed increase in variability associated with higher expected response levels. Moreover, the flexibility provided by the variance parameter $p$ enables the model to accommodate different degrees of risk escalation as the mean increases.

Finally, the presence of both the dispersion parameter $\phi$ and the weight parameter $w$ in the variance structure plays an important role, as it allows the model to capture additional sources of heterogeneity across observations.

\subsubsection{Loss Cost Variable} 

We now consider the stochastic counterpart of the observed loss cost introduced in Section~\ref{empiricalanalysis}. Let $N$ denote the number of claims reported for a given contract, and let $Z_k$ denote the cost of the $k$-th claim, for $k=1,\ldots,N$ when $N>0$. The loss cost, denoted by $Y$, is defined as
\[
Y=\begin{cases}
\sum_{k=1}^N Z_k, & \text{if } N>0,\\[0.2cm]
0, & \text{if } N=0.
\end{cases}
\]

Although each claim amount $Z_k$ is a continuous random variable, the aggregate quantity $Y$ is not itself continuous. In particular, the probability that no claim occurs ($N=0$) is strictly positive, implying that the distribution of $Y$ has a point mass at zero. The loss cost is therefore a semi-continuous random variable. 

This feature motivates the widespread use of the Tweedie distribution in insurance pricing models. In particular, when the loss cost $Y$ follows a Tweedie distribution with variance parameter satisfying \(1 < p < 2\), it admits a compound Poisson–Gamma representation, as discussed in \citep{tweedie1984index,jorgensen1997theory,delong2021making}. More precisely, the claim count $N$ follows a Poisson distribution, while the individual claim amounts $Z_k$ are assumed to be independent and gamma-distributed.

Under this representation, the Tweedie distribution can be interpreted as a mixed discrete–continuous distribution, making it a natural candidate for modeling loss cost $Y$. In addition, an exponential dispersion formulation can be derived, which facilitates statistical inference and premium modeling. The corresponding probability density function is given by

\begin{equation} \label{tweediedensityfunction}
f(y) = \exp\left(\frac{w}{\phi} \left(\frac{\mu^{\,1-p}}{1-p}\, y - \frac{\mu^{\,2-p}}{2-p} \right) + a(y,w,\phi)\right), \qquad y \in \mathbb{R},
\end{equation}
where \(a(\cdot)\) is a normalizing function that does not depend on \(\mu\).

Throughout the remainder of the paper, we assume that the variance parameter satisfies \(1 < p < 2\). For notational convenience, we write
\[
Y \sim \twee(\mu, w, \phi, p).
\]

\subsubsection{Weight Interpretation}

To better understand the role of the weight parameter in the Tweedie distribution, we recall its equivalence with the compound Poisson--Gamma representation, as summarized by \cite{delong2021making}. Suppose that a random variable \(Y\) follows a Tweedie distribution, that is, \(Y \sim \twee(\mu, w, \phi, p)\). Then \(Y\) can be represented as a compound Poisson--Gamma random variable with the following components:

\begin{itemize}
    \item The mean of the Poisson component is
    \[
        w\frac{\mu^{2-p}}{(2-p)\phi}.
    \]

    \item The shape parameter and the mean of the Gamma component are given respectively by
    \[
        \frac{2 - p}{p - 1}, 
        \qquad 
        \frac{(2 - p)\phi}{w \mu^{1 - p}}.
    \]
\end{itemize}

In this representation, the weight parameter \(w\) enters both components of the compound Poisson--Gamma distribution. It appears proportionally in the mean of the Poisson component, whereas it is inversely proportional to the mean of the Gamma component. This dual role is intuitive for the Poisson component, since \(w\) may naturally be interpreted as a measure of exposure or duration, consistent with the construction of a Poisson process.

In the special case where the weight parameter is set equal to the risk exposure, that is, \(w = t\), the resulting Tweedie model has limited practical appeal. In premium modeling, exposure \(t\) is typically incorporated within a Tweedie framework through either the offset approach or the ratio approach, as introduced earlier. Under both approaches, the premium $\mu$ is assumed to be proportional to the risk exposure $t$, in line with traditional pricing practice. 

The distinction between the two approaches lies in the specification of the weight parameter \(w\): the offset approach sets \(w = 1\), whereas the ratio approach specifies \(w = t^{\,p-1}\). The ratio formulation is arguably more intuitive, as exposure \(t\) enters exclusively through the Poisson component of the compound Poisson–Gamma representation. This is consistent with actuarial pricing principles, whereby exposure primarily affects claim frequency rather than claim severity.

\subsection{A Flexible Specification for Incorporating Risk Exposure}

As illustrated in the top panel of Figure~\ref{LCbyexposure}, the empirical relationship between exposure and observed loss cost does not follow the simple linear proportionality between premium and risk exposure assumed in traditional approaches. In particular, the average loss cost does not increase at a constant rate with respect to exposure, suggesting that the traditional specification may underestimate expected losses for contracts with low exposure. This departure from proportionality indicates that exposure does not act merely as a mechanical scaling factor, but instead interacts with the claim-generating process in a more intricate manner.

To account for this empirical behavior, we relax the proportionality constraint and allow exposure to affect the premium through a flexible adjustment function, as discussed in Section~\ref{Sect3}. Accordingly, the premium can be written in the additive form
\[
\mu = \exp\!\left\{\mathbf{x}^{\top}\boldsymbol{\beta} + \log\big(\gamma(t)\big)\right\},
\]

which makes explicit the smooth contribution of exposure through the term \(\log(\gamma(t))\).

To estimate \(\boldsymbol{\beta}\), we assume that the response variable follows a Tweedie distribution with weight parameter \(w\), while the dispersion parameter \(\phi\) and the variance power parameter \(p\) are treated as fixed constants. Under this specification, the estimation problem naturally falls within the framework of GAMs, as introduced in Section~\ref{Sect3}, with \(\log(\gamma(t))\) approximated using spline basis functions; see \cite{wood2017} for further details.

\subsubsection{Consistency Under the Traditional and Flexible Approaches}

The advantages of the flexible approach over the traditional one stem from the statistical properties of the corresponding estimators of the parameter vector \(\boldsymbol{\beta}\). We consider the same \(n\) independent loss costs as in Section~\ref{empiricalanalysis} and assume that they are generated from a common true distribution with density \(g(\cdot \mid \boldsymbol{\beta}^{\text{True}}, t)\), where \(\boldsymbol{\beta}^{\text{True}}\) denotes the true parameter vector. Although no specific distributional form is imposed on \(g\), we assume that the true expected premium is linked to the covariate vector \(\mathbf{x}\) and the risk exposure \(t\) through

\[
\mu^{\text{True}}
= \int_0^{+\infty} y\, g(y \mid \boldsymbol{\beta}, t)\, dy
= \delta(t)\,\exp\!\left\{\mathbf{x}^{\top}\boldsymbol{\beta}^{\text{True}}\right\},
\]

where \(\delta(t)\) is a positive function of \(t\). Our objective is to show that when traditional approaches are used to estimate \(\boldsymbol{\beta}^{\text{True}}\), the resulting estimator is consistent \emph{if and only if} 
\(\delta(t)=t\). In other words, consistency is obtained only when the true mean is proportional to the exposure, as assumed in traditional approaches.

We recall the definition of consistency from \cite{casella2024statistical}: an estimator \(\widehat{\boldsymbol{\beta}}\) is consistent for \(\boldsymbol{\beta}\) if

\[
\lim_{n\to\infty}
\Pr\!\left(|\widehat{\boldsymbol{\beta}}-\boldsymbol{\beta}| > \epsilon\right)
= 0
\qquad \text{for all } \epsilon>0.
\]

Because the true data-generating process \(g\) differs in general from the model assumed under traditional approaches, the model may be misspecified. We therefore rely on the theory of quasi--maximum likelihood estimation (or \emph{pseudo--maximum likelihood}), developed by \cite{white1982maximum}. Let \(f(\cdot \mid \boldsymbol{\beta}, t)\) denote the Tweedie density used in traditional approach, whose mean parameter is $\mu = t\,\exp\!\left(\mathbf{x}^{\top}\boldsymbol{\beta}\right)$.

The Kullback--Leibler divergence between the true density \(g\) and the assumed Tweedie density \(f\) is defined as

\[
KL(\boldsymbol{\beta})
= \int_0^{+\infty}
\log\!\left( \frac{g(y \mid \boldsymbol{\beta},t)}{f(y \mid \boldsymbol{\beta},t)} \right)
g(y \mid \boldsymbol{\beta},t)\,dy.
\]

According to \cite{white1982maximum}, the probability limit of \(\widehat{\boldsymbol{\beta}}\) is the minimizer of \(KL(\boldsymbol{\beta})\), which is equivalent to maximizing the expected log-likelihood:

\[
\int_0^{+\infty}\!\log f(y\mid\boldsymbol{\beta}, t)\, g(y\mid\boldsymbol{\beta}, t)\,dy
=
\int_0^{+\infty}
\left(
\frac{w}{\phi}
\left(
\frac{\mu^{1-p}}{1-p}y
- \frac{\mu^{2-p}}{2-p}
\right)
+ a(y,t,\phi)
\right)
g(y\mid\boldsymbol{\beta},t)\,dy,
\]

where \(w\) denotes any weight parameter used in the traditional approaches. In this setting, maximizing the expected log-likelihood is equivalent to maximizing the criterion

\begin{eqnarray}
K(\boldsymbol{\beta})
= \frac{\mu^{1-p}}{1-p}\,\mu^{\text{True}}
- \frac{\mu^{2-p}}{2-p}. \label{KappaBeta}
\end{eqnarray}

The gradient of \(K(\boldsymbol{\beta})\) with respect to \(\boldsymbol{\beta}\) is

\begin{eqnarray}
\nabla_{\boldsymbol{\beta}} K(\boldsymbol{\beta}) =\Big(\delta(t)\,t^{1-p} \exp\!\big\{(1-p)\mathbf{x}^{\top}\boldsymbol{\beta} + \mathbf{x}^{\top}\boldsymbol{\beta}^{\text{True}}\big\}- t^{2-p} \exp\!\big\{(2-p)\mathbf{x}^{\top}\boldsymbol{\beta}\big\} \Big)\mathbf{x}.
\label{Gradientnabla}
\end{eqnarray}

It follows that the true parameter vector \(\boldsymbol{\beta}^{\text{True}}\) satisfies the first-order condition \emph{if and only if} \(\delta(t)=t\). Consequently, estimators obtained from traditional approaches converge to the true parameter vector only when the true mean is proportional to the risk exposure, exactly as assumed in these methods. It is worth noting that this convergence does not depend on the weight, dispersion, or variance parameters specified under these traditional approaches.

Similarly, suppose that the function \(f\) corresponds to the Tweedie density used in the flexible approach, whose mean parameter is specified as
\[
\mu = \gamma(t)\,\exp\!\left(\mathbf{x}^{\top}\boldsymbol{\beta}\right).
\]

In this case, we obtain another maximization criterion of the form given in Equation~\ref{KappaBeta}, based on this alternative mean specification. The associated gradient of this criterion is
\[
\nabla_{\boldsymbol{\beta}} K(\boldsymbol{\beta})=\Big(
\delta(t)\,\gamma(t)^{\,1-p}\exp\!\big\{(1-p)\,\mathbf{x}^{\top}\boldsymbol{\beta} + \mathbf{x}^{\top}\boldsymbol{\beta}^{\text{True}}\big\}
-\gamma(t)^{\,2-p} \exp\!\big\{(2-p)\,\mathbf{x}^{\top}\boldsymbol{\beta}\big\}\Big)\mathbf{x}.
\]

It follows that the true parameter vector \(\boldsymbol{\beta}^{\text{True}}\) satisfies the first-order condition if and only if \(\delta(t)=\gamma(t)\). This result implies that, when the true mean loss cost is correctly specified—as assumed under the flexible approach—the corresponding estimator of the parameter vector is consistent. Determining which specification of the true mean is most appropriate can be guided by examining the empirical relationship between the average loss cost and the risk exposure, as we did in the data description section.

It is also important to note that, although the possibility that \(\delta(t)=t\) cannot be excluded, flexible methods reduce the risk of functional misspecification. In particular, the GAM specification adopted in the flexible approach enables the estimation procedure to capture, or closely approximate, the true functional relationship \(\delta(t)\) through spline-based approximations, regardless of whether \(\delta(t)\) coincides with \(t\).

\subsection{Choice of the Weight Parameter}

As shown above, the weight parameter $w$ of the underlying Tweedie distribution does not affect the consistency of the parameter vector estimator under the flexible specification. However, in order to obtain an estimate of this parameter vector, a specific value of $w$ must be supplied. A natural choice in an insurance pricing context is to let $w$ depend on the risk exposure $t$, that is, to set $w = \omega(t)$. In this regard, extending the approach proposed in \cite{raissaboucher2025} for a flexible function $\gamma(t)$, we consider two specifications for the weight function $\omega(\cdot)$ within the proposed framework:

\begin{enumerate}
    \item \textbf{Constant-Weight Model (CWM).} 
    This specification sets $\omega(t)=1$. It can be viewed as a generalization of the  \textit{offset approach}, where all observations receive the same weight regardless of the exposure duration.

    \item \textbf{Gamma-Weight Model (GWM).} 
    This specification sets $\omega(t)=\gamma(t)^{\,p-1}$, which generalizes the \textit{ratio approach}. This choice is motivated by the fact that the weight parameter appears exclusively in the Poisson component of the compound Poisson--Gamma representation of the Tweedie distribution, making $\omega(t)$ interpretable as a measure of exposure duration in the underlying Poisson process.
\end{enumerate}

We further assume that $\omega(1) = 1$. This assumption sets the one-year contract as the reference level, ensuring that policies with shorter or longer exposure durations are evaluated relative to the standard annual contract. Accordingly, we redefine the weight for a general exposure level \(t\) as

\[
\omega(t) = \left(\frac{\gamma(t)}{\gamma(1)}\right)^{\,p-1}.
\]

It is also important to note that the estimation of the parameter vector in the GWM is inherently more challenging than in the CWM. In the CWM, the weights are constant across contracts, and the only non-linear component to estimate is the spline-based function \(\gamma(t)\). In contrast, fitting a Tweedie GAM under the weighting scheme  
\(\omega(t) = \left(\gamma(t)/\gamma(1)\right)^{p-1}\)  
is non-standard because the same unknown function \(\gamma(t)\)—represented through a spline basis—appears both in the mean specification and in the weight structure. This creates an implicit dependence between the smoothing component and the estimation weights, requiring a tailored estimation strategy to ensure numerical stability and convergence.

To address this difficulty, we adopt the following iterative estimation scheme:

\begin{itemize}
    \item \textbf{Initialization.}  
Fit an initial Tweedie GAM with mean  $\mu = s_1(t)\exp\{\mathbf{x}^{\top}\bm{\beta}\}$, and weights \(\omega(t)=t\), where \(s_1(t)\) is an initial spline approximation of \(\gamma(t)\).

\item \textbf{Iteration.}  
Given the spline estimate \(s_k(t)\) at iteration \(k\), fit a model with  $\mu = s_{k+1}(t)\exp\{\mathbf{x}^{\top}\beta\}$, and weights \(\omega = s_k(t)\), normalized so that \(\omega(1)=1\). 

\item \textbf{Convergence.}  
Repeat this procedure until the sequence \(s_k(t)\) stabilizes, yielding the final estimate of \(\gamma(t)\) and the corresponding regression parameter vector \(\boldsymbol{\beta}\).
\end{itemize}

This iterative algorithm ensures that the spline approximation of \(\gamma(t)\) is coherently incorporated into both the mean specification and the weight structure, thereby enabling a stable and internally consistent estimation of the Gamma-Weight Model.

We also argue that a third weighting structure deserves consideration. To motivate this proposal, we recall the ratio approach of \cite{raissaboucher2025}. As shown by the authors, this framework can be equivalently formulated by considering the aggregate loss cost \(Y\) as the response variable and assuming that it follows a Tweedie distribution whose mean parameter is proportional to the risk exposure \(t\), and whose weight parameter is given by \(t^{\,p-1}\).

The proposed third weighting structure then generalizes this approach by retaining the same weighting scheme \(\omega(t)=t^{\,p-1}\), while adopting a flexible specification for the mean of \(Y\). In this interpretation, we obtain a closed-form weighting function \(\omega(t)\) that is non-constant and enjoys practical appeal, as it is motivated by the same exposure-based weighting principle underlying the ratio approach. In line with the two models introduced above, we therefore define this third specification as follows:

\begin{enumerate}
\setcounter{enumi}{2}
\item \textbf{Exposure-Weighted Model (EWM).} A generalization of the ratio-based approach obtained by specifying \(\omega(t)=t^{\,p-1}\) while allowing for a flexible exposure effect in the mean structure.
\end{enumerate}

\subsection{Model Selection}

Model comparison in actuarial science is commonly performed using scoring rules derived from the predictive distribution of risk, as discussed in \cite{scoringrules}. Among these, the logarithmic score is particularly appealing. However, the log-likelihood of the Tweedie distribution does not admit a closed-form expression, which restricts its direct application. For this reason, we rely on a deviance-based model selection criterion to guide the comparison among the three proposed models as well as the traditional approaches.

It should be emphasized, however, that the deviance-based selection criterion is purely statistical in nature, as recalled below. For this reason, we also consider complementary evaluation measures. In particular, we examine the ABC (Area Between the Curves) criterion, derived from concentration and Lorenz curves, as summarized in \cite{denuit2019model}, and a criterion based on Murphy diagrams, which relies on Bregman dominance, as presented in \citet{denuit2025comparison}.

\subsubsection{Assessing Model Fit and Comparison Using the Deviance}

We consider the same \(n\) independent loss costs \(y_1,\ldots,y_n\) as previously introduced. We denote by \(\mu_i\) the corresponding premium for observation \(i\), and by \(t_i\) its risk exposure. According to \cite{nelder1972generalized}, the deviance of a regression model is defined as follows:

\[
\mathcal{D}(\widehat{\boldsymbol{\beta}}) 
= 2\phi\left(\ell^{*} - \ell(\widehat{\boldsymbol{\beta}})\right),
\]
where $\ell^{*}$ denotes the log-likelihood of the saturated model, and $\ell(\widehat{\boldsymbol{\beta}})$ is the log-likelihood evaluated at the estimated model. For the Tweedie model, the deviance takes the explicit form
\[
\mathcal{D}(\widehat{\boldsymbol{\beta}})
=
2 \sum_{i=1}^n 
\omega(t_i)
\left[
y_i\left(\frac{y_i^{\,1-p} - \widehat{\mu}_i^{\,1-p}}{1-p}\right)
-
\frac{y_i^{\,2-p} - \widehat{\mu}_i^{\,2-p}}{2-p}
\right],
\]

where \(\omega(t_i)\) is replaced by the weight used in the corresponding model, and \(\widehat{\mu}_i\) denotes the premium estimate obtained under that same model. The deviance provides a quantitative measure of lack of fit: smaller values indicate a model that better captures the structure of the data. Consequently, when comparing two competing models, the preferred model is the one with the lower deviance. In the numerical application, we explicitly perform this comparison by splitting the dataset into training and test samples, and by computing the deviance on both samples. The purpose of this split is to assess whether the competing models exhibit overfitting, which would typically manifest itself through  a substantial increase in the deviance when moving from the training sample to the test sample. Ideally, one would expect only a small gap between the deviances computed on the two samples.

\subsubsection{Model Comparison Using Concentration and Lorenz Curves} \label{abc.criterion}

To compare competing premium estimators, we use the concentration curve, the Lorenz curve, and the associated ABC; see \cite{denuit2019model}. Let $\mu_i$ denote the true premium for contract $i$, let $\widehat{\mu}_i$ be a candidate premium estimator, and let $y_i$ be the observed loss cost. We write $\mu$, $\widehat{\mu}$ and $Y$ for the corresponding generic random variables, and assume that $\Esp{\mu}<+\infty$, $\Esp{\widehat{\mu}}<+\infty$ and $\Esp{Y}<+\infty$.

Let $F$ denote the distribution function of $\widehat{\mu}$, that is,
\[
F(x)=\Prr{\widehat{\mu}\le x}, \qquad x\in\mathbf{R},
\]
and let
\[
F^{-1}(\theta)=\inf\{x\in\mathbf{R}:F(x)\ge \theta\}, \qquad 0\le \theta \le 1,
\]
be its generalized inverse. The concentration curve is defined by
\begin{equation}\label{cc.curve.1}
CC[\mu,\widehat{\mu};\theta]=\frac{\Esp{\mu\,\mathbf{1}_{\{\widehat{\mu}\le F^{-1}(\theta)\}}}}{\Esp{\mu}},
\qquad 0\le \theta \le 1.
\end{equation}

As shown in \cite{denuit2019model}, it can equivalently be written as

\begin{equation}\label{cc.curve.2}
CC[Y,\widehat{\mu};\theta] = \frac{\Esp{Y\,\mathbf{1}_{\{\widehat{\mu}\le F^{-1}(\theta)\}}}}{\Esp{Y}},
\qquad 0\le \theta \le 1,
\end{equation}
which is the formulation used in practice since $\mu$ is not observable.

The Lorenz curve associated with $\widehat{\mu}$ is
\begin{equation}\label{lc.curve}
LC[\widehat{\mu};\theta]
=
\frac{\Esp{\widehat{\mu}\,\mathbf{1}_{\{\widehat{\mu}\le F^{-1}(\theta)\}}}}{\Esp{\widehat{\mu}}},
\qquad 0\le \theta \le 1.
\end{equation}
The ABC criterion is then defined as
\[
ABC(\widehat{\mu})=\int_0^1\Big(CC[Y,\widehat{\mu};\theta]
-LC[\widehat{\mu};\theta]\Big)\,d\theta.
\]

Following \cite{denuit2024convex,denuit2024testing}, the comparison of competing premium estimators through the ABC criterion is meaningful once they are put on a common globally balanced scale. For this reason, a candidate estimator $\widehat{\mu}$ may first be rescaled as
\[
\widehat{\mu}_c= \frac{\Esp{Y}}{\Esp{\widehat{\mu}}}\,\widehat{\mu},
\]
so that $\Esp{\widehat{\mu}_c}=\Esp{Y}$. Since this transformation preserves the ranking induced by $\widehat{\mu}$, it does not modify the concentration curve. The competing estimators can then be compared through their ABC values, smaller values indicating a closer proximity to local balance. 

\subsubsection{Murphy Diagrams}

We adopt the same notation as in the previous subsection. In particular, $\mu$ denotes the true premium, $Y$ the loss cost, and $\widehat{\mu}^{(1)}$ and $\widehat{\mu}^{(2)}$ two competing premium estimators. Following \citet{denuit2025comparison}, we say that $\widehat{\mu}^{(2)}$ Bregman-dominates $\widehat{\mu}^{(1)}$ if
\[
\Esp{L(Y,\widehat{\mu}^{(2)})}
\;\le\;
\Esp{L(Y,\widehat{\mu}^{(1)})}
\]
for every Bregman loss function $L$. This dominance relation provides a comparison criterion that does not depend on the choice of a particular loss function within the Bregman class.

Such a comparison can be examined graphically by means of Murphy diagrams. To this end, consider the elementary loss functions
\[
L_m(Y,\widehat{\mu}^{(k)})= \frac{1}{2}(m-Y)\,\mathbf{1}_{\{\widehat{\mu}^{(k)}>m\}},
\qquad k=1,2,
\]
indexed by $m\in\mathbf{R}_+$. For each estimator $\widehat{\mu}^{(k)}$, the Murphy diagram is the graph of
\[
m \longmapsto \Esp{L_m(Y,\widehat{\mu}^{(k)})},
\qquad m\in\mathbf{R}_+.
\]

If
\[
\Esp{L_m(Y,\widehat{\mu}^{(2)})}
\;\le\;
\Esp{L_m(Y,\widehat{\mu}^{(1)})}
\qquad \text{for all } m\in\mathbf{R}_+,
\]

then $\widehat{\mu}^{(2)}$ dominates $\widehat{\mu}^{(1)}$ in the sense of Murphy diagrams, and therefore Bregman-dominates it; see \citet{denuit2025comparison}. Murphy diagrams thus provide a convenient graphical tool for comparing premium estimators over the whole class of Bregman losses. In contrast with the $ABC$ criterion, which evaluates proximity to local balance once global balance has been enforced, Murphy diagrams assess dominance directly through expected loss.

\section{Numerical Application}

\label{sectNum}

To illustrate the proposed modeling framework, we apply the different Tweedie specifications to the dataset introduced in Section~\ref{SectData}. Recall that this dataset consists of a non-random sample extracted from a Canadian automobile insurance portfolio.  It includes contracts of type $\mathcal{XX}$ (fully exposed to risk for the entire policy period) and $\mathcal{XO}$ (initially exposed but canceled before expiration). In addition to the covariates described in Section~\ref{SectData}, the BMS level is included as an additional rating factor, as discussed in Section~\ref{SectDataBMS}.

\subsection{Adjusting the Models Under Different Weighting Schemes}

We examine the three proposed flexible specifications—CWM, GWM, and EWM—and compare them with the traditional ratio approach, which has been recommended as superior to the offset approach for incorporating risk exposure $t$ in conventional actuarial models \citep{raissaboucher2025}.

For the GWM, Figure~\ref{Appendix} in Appendix illustrates the evolution of the weights at each iteration $k$ of our algorithm, providing an informal visualization of its convergence. Convergence is extremely rapid: after only two or three iterations, the sequence of weights already appears to have reached its final shape.

All three models are calibrated using the training dataset, while the test dataset is reserved for evaluating their predictive performance. We do not estimate the dispersion parameter $\phi$ and instead fix the variance parameter at $p = 1.42$, following the estimate obtained in \cite{raissaboucher2024}. To ensure a fair comparison across approaches, we use a cubic regression spline with the same set of knots. Although the choice of spline type does not affect our conclusions, enforcing the same knots and basis functions across models facilitates meaningful comparisons.

\subsubsection{Weights and Spline-Based Estimation Analysis}

\begin{figure}[H]
    \begin{center}
        \includegraphics[scale=0.54]{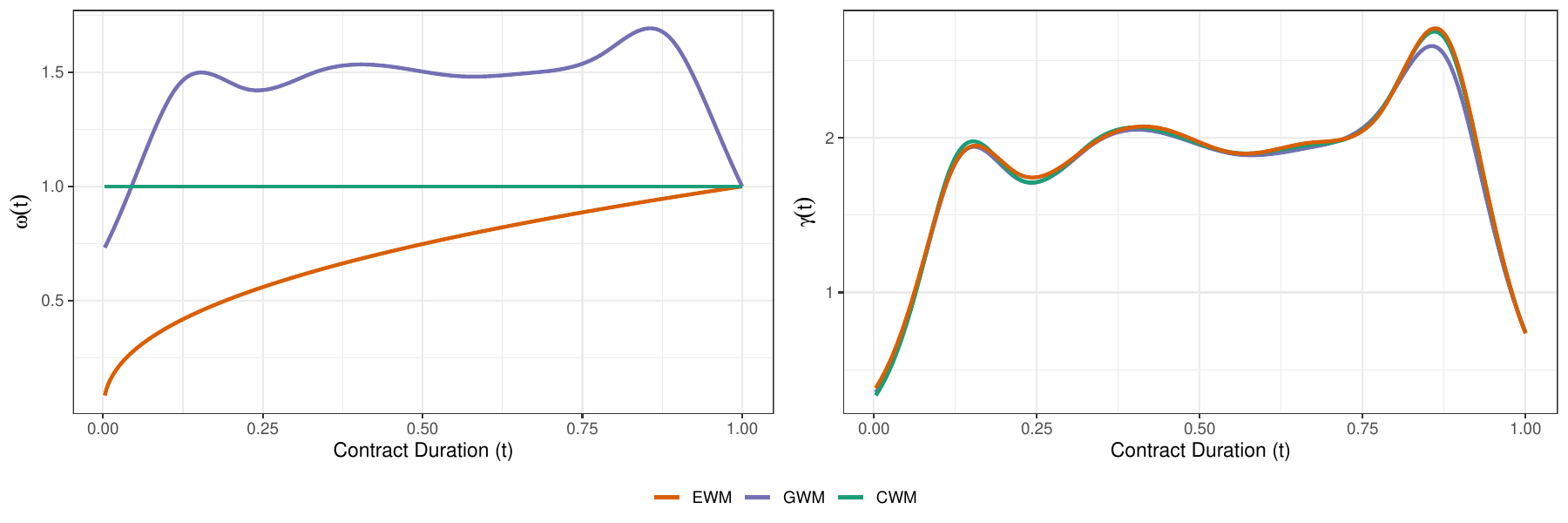}
        \caption{Weight function $\omega(t)$ (left) and spline-based estimation of $\gamma(t)$ (right) for the flexible approaches}
        \label{w1}
    \end{center}
\end{figure}

The left panel of Figure~\ref{w1} displays the weight functions $\omega(t)$ obtained for the three models under study. In the CWM, all observations contribute equally to the estimation of the Tweedie model parameters. In other words, each policy has the same influence on the estimation process, even if it has been in force for only a short period—sometimes just a few weeks or even days. In contrast, the two other approaches, namely the GWM and the EWM, assign different weights to observations with exposures lower than one. The EWM behaves intuitively, assigning increasing weights as exposure increases. The GWM, however, exhibits a more peculiar pattern: it assigns relatively higher weights to policies that were canceled mid-term.

Although the exposure $t$ does not appear explicitly in the weight function $\omega(t)$ under the CWM, it is still incorporated into the model through the exposure function $\gamma(t)$, which is estimated using spline basis functions and shown in the right panel of Figure~\ref{w1}. This figure reveals that, although the estimated exposure curves have similar shapes across the three models, small but noticeable differences remain. For instance, the curves associated with the CWM and EWM tend to lie slightly below the one obtained under the GWM. 

\subsubsection{Mean Parameter Adequacy Test}

\begin{figure}[H]
	\begin{center}
		\includegraphics[scale=0.54]{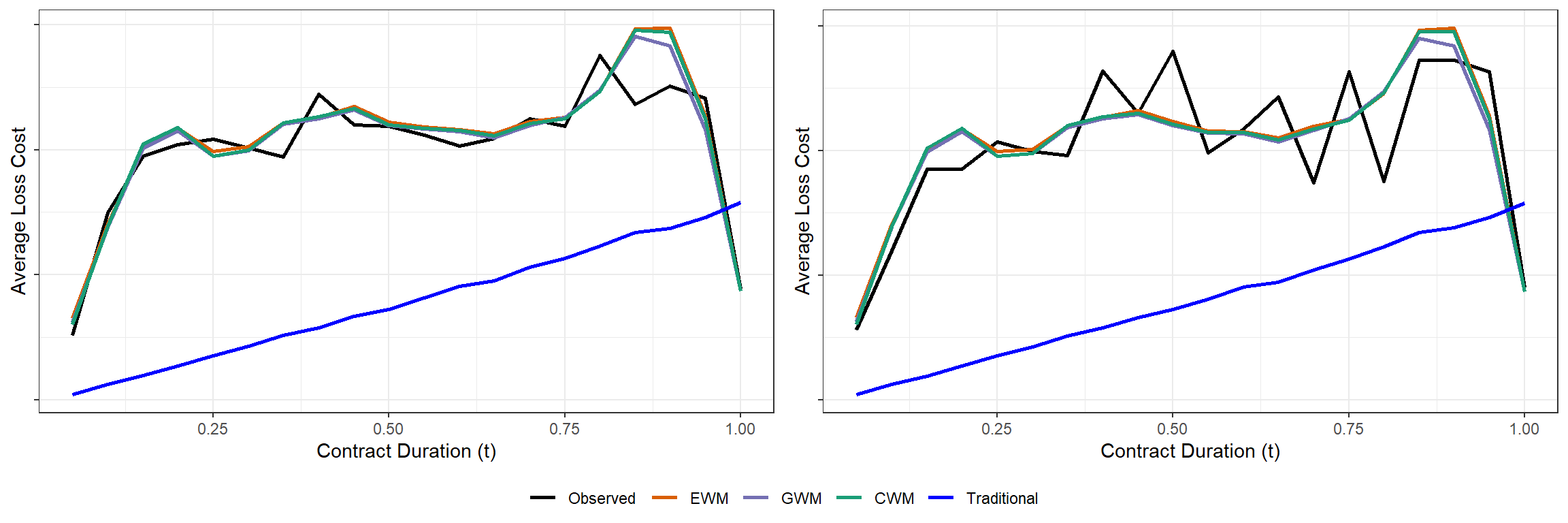}
	\caption{Observed average loss costs and estimated average premiums (left: training set, right: test set)}
		\label{mucomparison}
	\end{center}
\end{figure}

We verify our assumption regarding the mean structure in the proposed flexible approaches by comparing the average loss cost and the average estimated premiums in both the training and test datasets, as illustrated in Figure~\ref{mucomparison}. Before commenting on this figure, we examine the estimated coefficients reported in Table~\ref{tab:beta_schemes}. This table presents the estimated $\beta$ parameters included in the mean structure, corresponding to the five segmentation covariates as well as the effect of the BMS level. It is interesting to observe that the estimated $\beta$ coefficients exhibit more noticeable variation across models than the spline basis coefficients. This suggests that the weighting scheme primarily affects the relative importance of the covariates rather than the overall shape of the exposure adjustment.

\begin{table}[H]
   \centering
   \begin{tabular}{l|c|ccc}
\hline
Parameters & Traditional approach & \multicolumn{3}{c}{Flexible approach} \\
           &                      & CWM & GWM & EWM \\
\hline
$\beta_1$             & -0.158 & -0.216 & -0.235 & -0.185 \\
$\beta_2$             &  0.232 &  0.162 &  0.153 &  0.170 \\
$\beta_3$             &  0.728 &  0.878 &  0.839 &  0.926 \\
$\beta_4$             &  0.252 &  0.360 &  0.340 &  0.380 \\
$\beta_5$             & -0.239 & -0.145 & -0.158 & -0.131 \\
$\beta_{\text{BMS}}$  &  0.130 &  0.108 &  0.108 &  0.107 \\
\hline
\end{tabular}
   \caption{Estimated parameter vectors for all models}
   \label{tab:beta_schemes}
\end{table}

The mean specification assumed under the flexible approach appears to provide an adequate fit to the loss-cost data, as illustrated in Figure~\ref{mucomparison}. This adequacy is observed for both the training and test datasets. We also note that the ratio approach—represented by the blue curve —which assumes proportionality between the premium and the risk exposure, appears to fit less well as the three flexible models, even though its increasing trend with respect to exposure is consistent with theoretical expectations.

\subsubsection{Analysis of Alternative Weight Selection Criteria}

As mentioned earlier, we rely on two additional criteria for selecting the weight specification: the deviance-based criterion (evaluated on both the training and test datasets) and the criterion derived from concentration and Lorenz curves.

For the deviance criterion, since the four models under comparison do not share the same weight function \(\omega(t)\)—although they use the same response variable—we introduce normalized observation weights. Let \(t_1,\ldots,t_n\) denote the risk exposures associated with the \(n\) observations in the dataset on which the deviance is evaluated. The normalized weight associated with observation \(i\) is defined as
\[
w_i = \frac{\omega(t_i)}{\sum_{j=1}^n \omega(t_j)}, \qquad i=1,\ldots,n.
\]

The advantage of this normalization is that it places all models on a comparable scale, as the scale effect of the original weight function is removed. Based on the normalized deviances in Table \ref{scoredev}, we observe that the flexible approaches outperform the traditional ratio approach, yielding lower deviance scores for both the training and test datasets. Among the flexible specifications, EWM performs best, achieving the lowest deviance values on both datasets. We can also observe that the deviance of the test dataset is comparable to that of the training dataset, which suggests that the proposed flexible framework does not suffer from overfitting.

\begin{table}[H]
\centering
\begin{tabular}{l|c|ccc}
\hline
Datasets & Traditional approach & \multicolumn{3}{c}{Flexible approach} \\
         &                      & CWM & GWM & EWM \\
\hline
Training set & 114.64 & 102.57 & 102.60 & 102.56 \\ 
Test set     & 114.77 & 102.77 & 102.80 & 102.76 \\ 
\hline
\end{tabular}
\caption{Deviance comparison across models (training and test sets)}
\label{scoredev}
\end{table}

\begin{figure}[H]
	\begin{center}
		\includegraphics[scale=0.54]{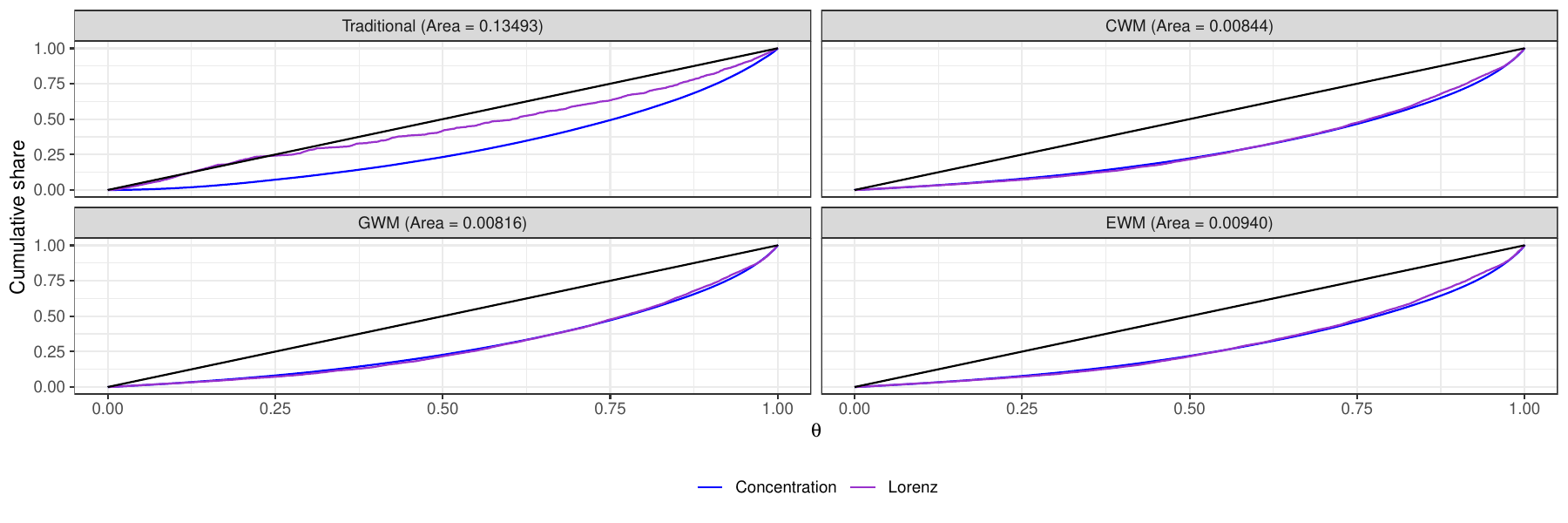}
  \caption{Concentration and Lorenz curves}
  \label{ccandlccurves}
	\end{center}
\end{figure}

Figure~\ref{ccandlccurves} is in line with the conclusions drawn from the deviance comparison. To construct this figure, we use the empirical counterparts of the concentration and Lorenz curves, computed on the training set; see \citet{denuit2019model}. For \(0\le \theta \le 1\), these are defined by
\[
\widehat{CC}(\theta)
=\frac{\sum_{i:\,\widehat{\mu}_i<\widehat{F}^{-1}(\theta)} y_i
}{\sum_{i=1}^n y_i},
\qquad
\widehat{LC}(\theta)=\frac{ \sum_{i:\,\widehat{\mu}_i<\widehat{F}^{-1}(\theta)} \widehat{\mu}_i}{\sum_{i=1}^n \widehat{\mu}_i},
\]

where \(y_i\) and \(\widehat{\mu}_i\) denote the observed values of \(Y\) and of the candidate premium estimator \(\widehat{\mu}\), respectively, and where \(\widehat{F}^{-1}(\theta)\) is the empirical quantile of order \(\theta\) computed from the empirical distribution of \(\widehat{\mu}\).

We then rescale \(\widehat{\mu}\) for all competing models, as described in Section~\ref{abc.criterion}. This rescaling ensures that global balance is achieved for each model. We subsequently compute the actual area between the curves using the following expression:
\[
\text{Area} = \int_0^1 \left| \widehat{CC}(\theta) - \widehat{LC}(\theta) \right| \, d\theta.
\]

In the present setting, the Area measure is preferred to the ABC criterion because the empirical concentration and Lorenz curves seem to intersect. In such circumstances, the ABC criterion is less informative, since it depends on the signed difference between the two curves and may therefore be subject to compensations between positive and negative discrepancies. Consequently, a small, or even null, ABC value need not reflect a genuine proximity between the curves, but may simply result from offsetting deviations. The Area measure is therefore more suitable here, as it provides a more faithful quantification of the discrepancy between the two curves.

The curves obtained from the three flexible approaches are nearly indistinguishable graphically (see Figure~\ref{ccandlccurves}). The Area measure shown in Figure~\ref{ccandlccurves} yields a substantially larger value for the traditional ratio approach (Area \(\approx 0.13\)) than for the flexible specifications. Among the latter, the Area values are very close to one another. Although the differences are minor, GWM produces the smallest Area, suggesting a slight advantage in terms of local balance.

\begin{figure}[H]
	\begin{center}
		\includegraphics[scale=0.54]{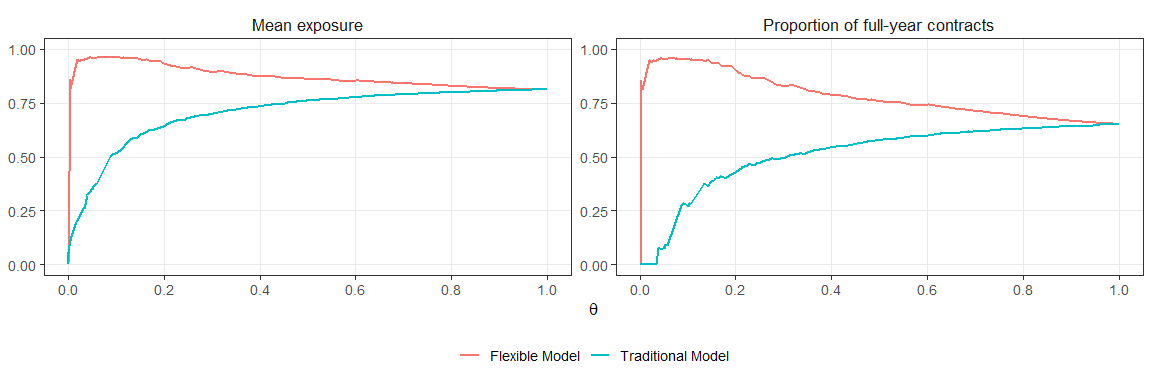}
\caption{Mean exposure and proportion of full-year contracts}
  \label{ccandlccurves_expo}
	\end{center}
\end{figure}

The strong performance of the flexible approaches is further clarified by Figure~\ref{ccandlccurves_expo}, which displays the mean exposure together with the proportion of full-year contracts along the same ordering induced by the estimated premiums. A marked contrast appears between the traditional and flexible specifications. For the flexible approach (illustrated here for the EWM), contracts with the lowest estimated premiums tend to have an average exposure close to \(1\), and thus correspond predominantly to policies running to maturity. By contrast, the traditional specification imposes a predetermined relationship between the premium and the exposure \(t\), and therefore cannot fully adjust the premium level for contracts that terminate before one year. As a result, short-exposure contracts tend to be underpriced relative to the flexible models, which helps explain the weaker overall performance of the traditional approach.

\subsection{Penalties and Ratemaking Strategies}
\label{SectPenRate}

The right-hand panel of Figure~\ref{w1} displays the function $\gamma(t)$ as a function of exposure $t$. With respect to the ratemaking constraints introduced in Section~\ref{sectPenalty}, it is clear that the resulting pattern is not practically applicable, even though the three models with a flexible specification of the mean outperform the traditional models on both the training and test datasets. In particular, under this specification, a policyholder may face a higher mid-term cancellation penalty when canceling after only $0.2$ years of coverage than if the policy were not canceled at all, which occurs when $\gamma(t)$ exceeds $1$. To obtain a more realistic and operational structure, it is therefore preferable to impose the ratemaking constraints directly on the spline basis functions used in the three flexible Tweedie models.

For illustrative purposes, however, we propose to transform the function $\gamma(t)$ into a new function $\gamma_{\text{con}}(t)$ that incorporates the desired ratemaking constraints, and to re-estimate all model parameters using this transformed function. This approach allows all parameters---in particular the regression coefficients $\beta$---to be estimated simultaneously, thereby enabling a partial adjustment to the imposed constraints. It should be noted that imposing structural constraints on $\gamma(t)$ modifies the estimation framework considered in Section~4. In particular, once monotonicity and ratemaking requirements are enforced, the estimator no longer arises from the unrestricted quasi--maximum likelihood setting of \cite{white1982maximum}, and consistency is therefore no longer guaranteed; this reflects a trade-off between statistical optimality and practical implementability. In practice, however, the imposed constraints are mild and primarily affect the shape of the penalty function rather than the exposure adjustment in the interior of the domain. As a result, the core exposure–risk relationship captured by $\gamma(t)$ remains largely preserved.

\begin{figure}[H]
	\begin{center}
		\includegraphics[scale=0.54]{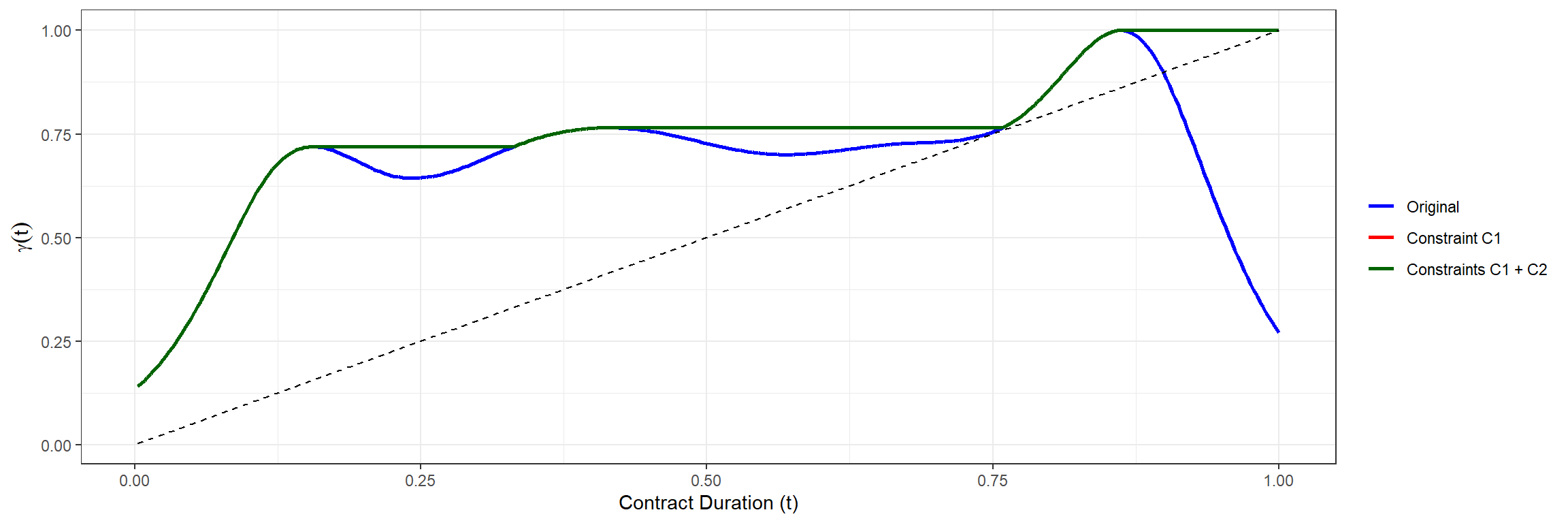}
	    \caption{Optimal penalty function as a function of exposure $t$}
		\label{penalty1}
	\end{center}
\end{figure}

Based on the results obtained under the Exposure-Weighted approach, Figure~\ref{penalty1} displays the optimal penalty as a function of the risk exposure $t$. According to our definition, the penalty function $\rho(\cdot)$ corresponds to the area between the diagonal dashed line and the green curve in Figure~\ref{penalty1}. The green curve represents the transformed function $\gamma_{\text{con}}(t)$ obtained by imposing both ratemaking constraints (C1) and (C2). In our numerical application, imposing constraint (C2) has only a negligible effect on the results, as illustrated in the figure.

It should be emphasized, however, that imposing $\gamma_{\text{con}}(t)$ to be increasing in the exposure $t$ and to lie everywhere above the identity function ($t \mapsto t$) does not guarantee that the penalty function $\rho(\cdot)$ is itself monotone in~$t$. As illustrated in Figure~\ref{penalty1}, it may occur that a contract canceled after nine months incurs a smaller penalty than one canceled earlier in the year.
Nevertheless, once the additional structural constraints of our pricing framework are imposed, we ensure that the \textit{total premium} paid by a policyholder canceling after nine months cannot fall below that of policyholders who cancel earlier in the year.

\begin{figure}[H]
	\begin{center}
		\includegraphics[scale=0.54]{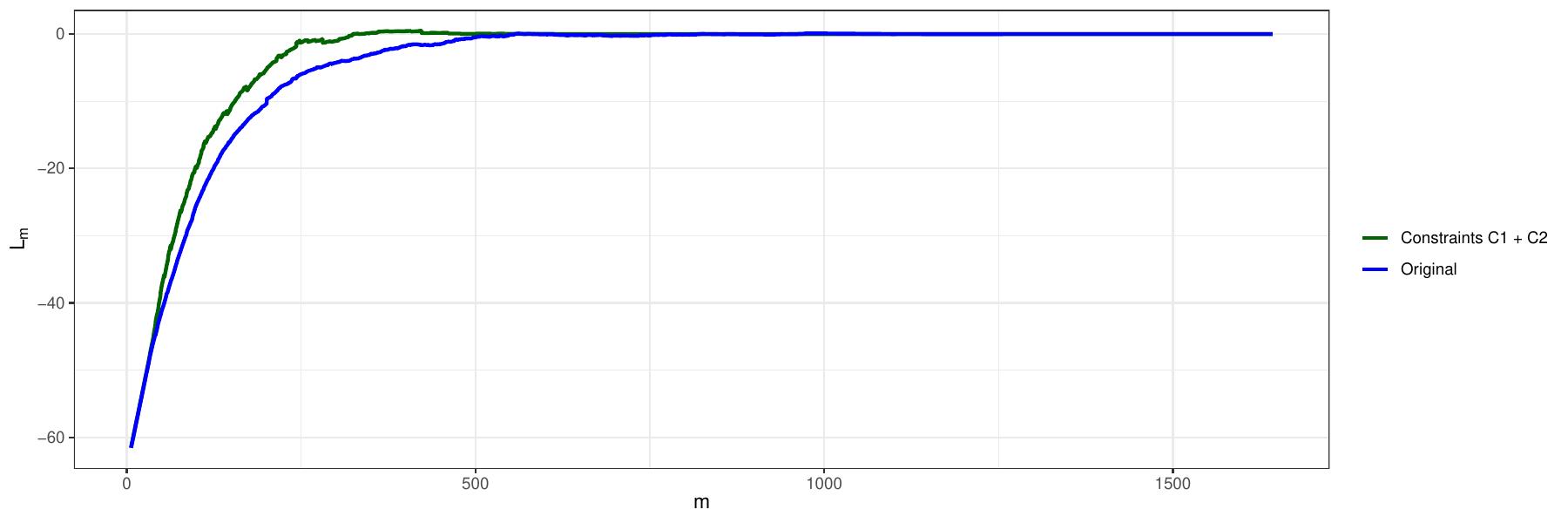}
	    \caption{Murphy diagram of the optimal penalty }
		\label{murphy}
	\end{center}
\end{figure}

We compare the performance of the model using $\gamma_{\text{con}}(t)$, obtained by imposing both ratemaking constraints (C1) and (C2), with the Exposure-weight approach, represented by the blue (original) line in Figure~\ref{murphy}. This figure displays the Murphy diagrams of the two models based on the following empirical version of the loss function:

\[
L_m = 
\frac{1}{2n} \sum_{i=1}^n (m - y_i)\,\mathbf{1}_{\{\widehat{\mu}_i > m\}}.
\]

We observe that the Exposure-weight model outperforms the constrained model for all values of $m$, which is consistent with our theoretical findings. Indeed, we previously established that the parameter vector estimator is consistent under the Exposure-weight specification, ensuring that it captures the same structural relationship as the true premium.

\subsubsection{Smoothing the Penalties}
\label{SectSmoothPen}

We observe that imposing constraint (C1) still produces extreme outcomes. After only a few weeks of insurance coverage, a policyholder would already be required to pay more than half (almost 75\%) of the full annual premium if they cancel their policy. Although this may appear unfair at first glance, it is important to emphasize that, based solely on the empirical results and without imposing constraint (C1), the situation would be even worse: the policyholder would have to pay more than the full annual premium in order to cancel mid-term.

It is worth noting that, as with many other products involving annual subscriptions or, more generally, continuous usage over time, it may not be unreasonable for the insurance industry to reconsider its current commercial practices and reflect on the possibility of requiring a minimum premium, even when coverage is canceled after only a few days. Although such an approach could eventually be proposed, our purpose here is not to recommend a new pricing policy. Instead, our objective is to introduce a correction to the penalty function obtained earlier, in order to produce a more realistic premium structure. To that end, we propose the following functional form:
\[
\gamma_{\text{adj}}(t, a) = a\, \gamma_{\text{con}}(t) + (1-a)\, t, \qquad 0 \le a \le 1.
\]

\begin{figure}[H]
	\begin{center}
		\includegraphics[scale=0.54]{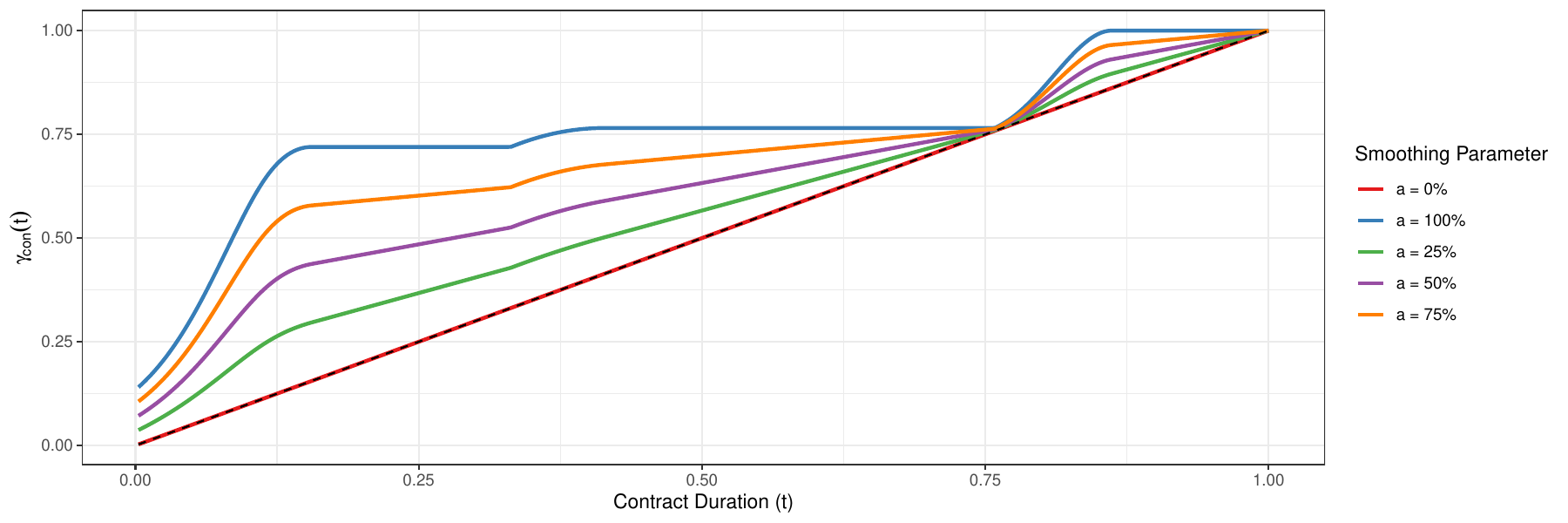}
		\caption{Effect of the adjustment parameter $a$ on the penalty functions}
		\label{penalty2}
	\end{center}
\end{figure}

Figure~\ref{penalty2} illustrates several values of $a$ and their corresponding effects on the penalty function. As can be seen, setting $a = 100\%$ recovers the optimal penalty function previously discussed, whereas $a = 0\%$ corresponds to a null penalty. This figure also highlights that the penalty increases as $a$ increases. Hence, insurers could use the parameter $a$ to adopt a more or less stringent penalty for mid-term cancellations.
 
\subsubsection{Tweedie Model with an Imposed Penalty Structure}

It then becomes possible to apply a Tweedie model under constraints (C1) and (C2), together with the various penalty shapes proposed through $\gamma_{\text{adj}}(t,a)$. This is achieved by directly incorporating the function $\log\!\left(\gamma_{\text{adj}}(t,a)\right)$ as an offset term in the mean structure of the Tweedie model. Formally, the mean parameter is specified as $ \mu = \exp\!\left( \mathbf{x}^\top \boldsymbol{\beta} \;+\; \log\!\left(\gamma_{\text{adj}}(t,a)\right) \right)$, where $\boldsymbol{\beta} = (\beta_0, \beta_1, \ldots, \beta_5, \beta_{\text{BMS}})^\top$ denotes the parameter vector.

\begin{figure}
    \centering
    \begin{subfigure}{0.32\textwidth}
        \centering
        \includegraphics[width=\linewidth]{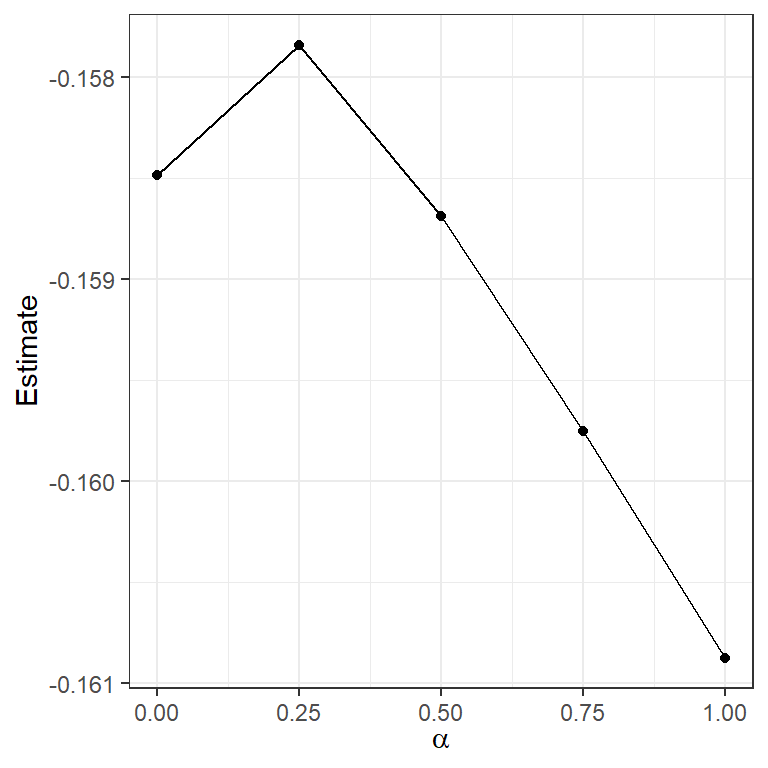}
        \caption{$\beta_1$}
    \end{subfigure}
    \begin{subfigure}{0.32\textwidth}
        \centering
        \includegraphics[width=\linewidth]{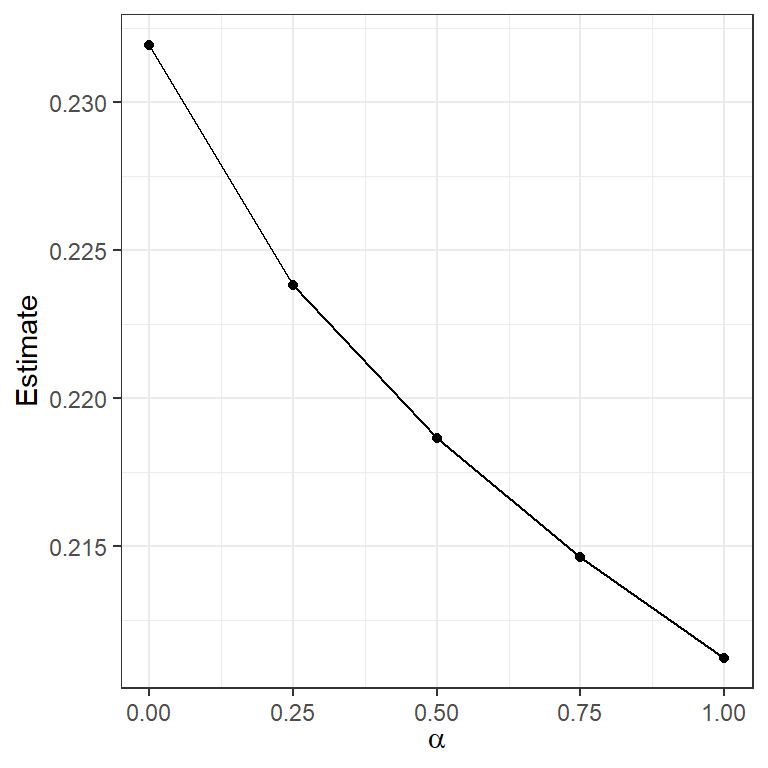}
        \caption{$\beta_2$}
    \end{subfigure}
    \begin{subfigure}{0.32\textwidth}
        \centering
        \includegraphics[width=\linewidth]{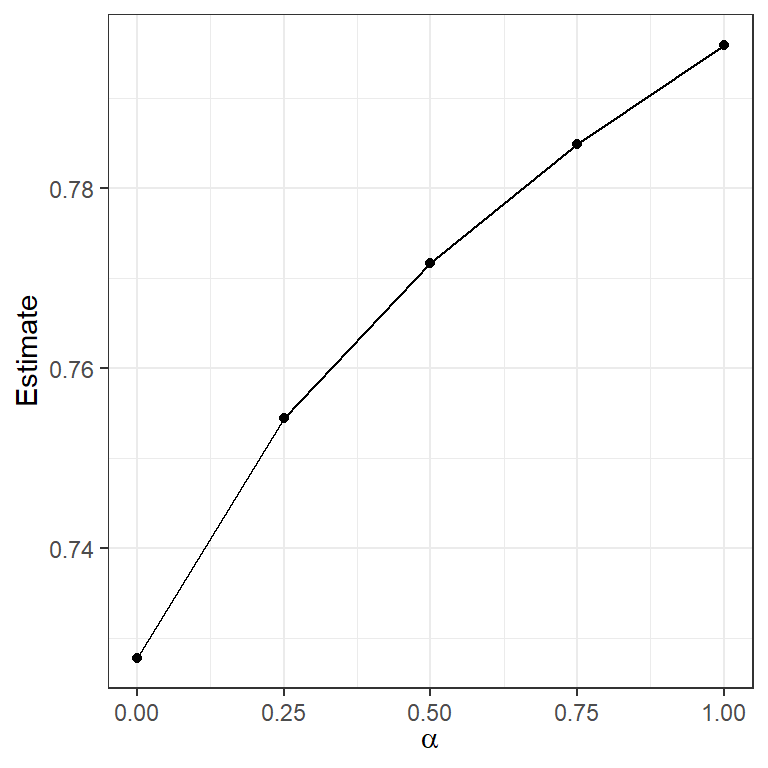}
        \caption{$\beta_3$}
    \end{subfigure}
    
    \begin{subfigure}{0.32\textwidth}
        \centering
        \includegraphics[width=\linewidth]{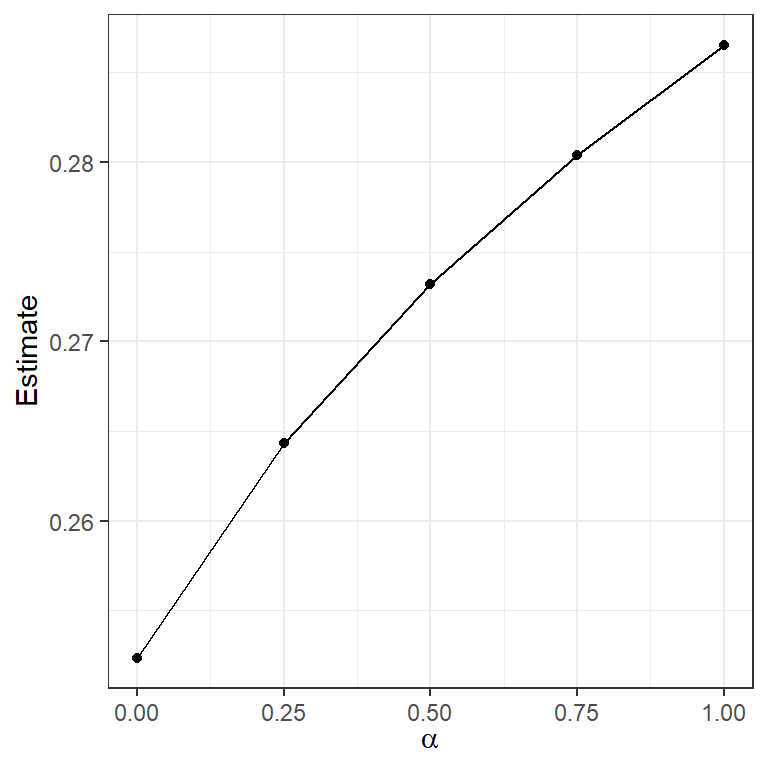}
        \caption{$\beta_4$}
    \end{subfigure}
    \begin{subfigure}{0.32\textwidth}
        \centering
        \includegraphics[width=\linewidth]{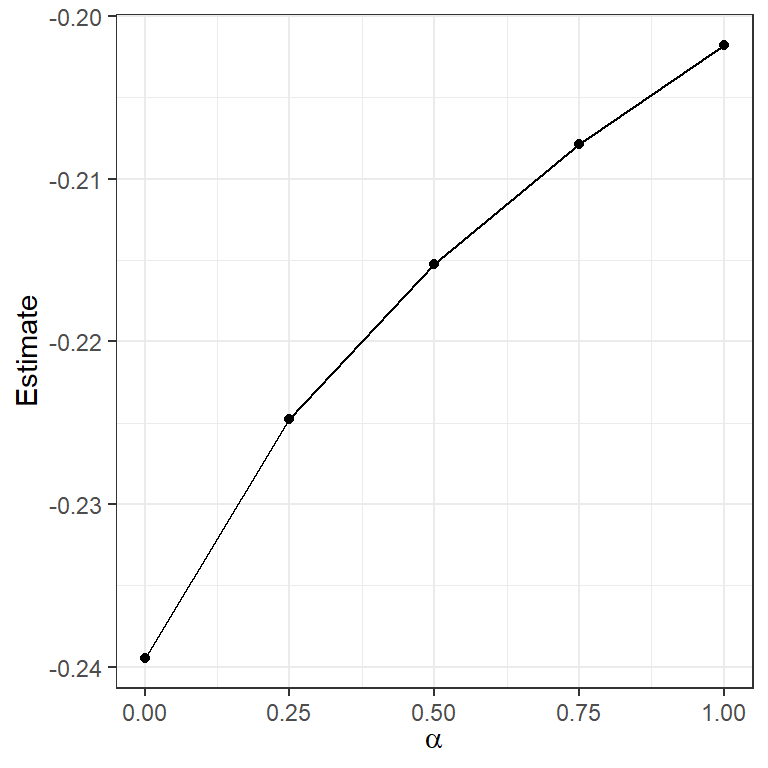}
        \caption{$\beta_5$}
    \end{subfigure}
    \begin{subfigure}{0.32\textwidth}
        \centering
        \includegraphics[width=\linewidth]{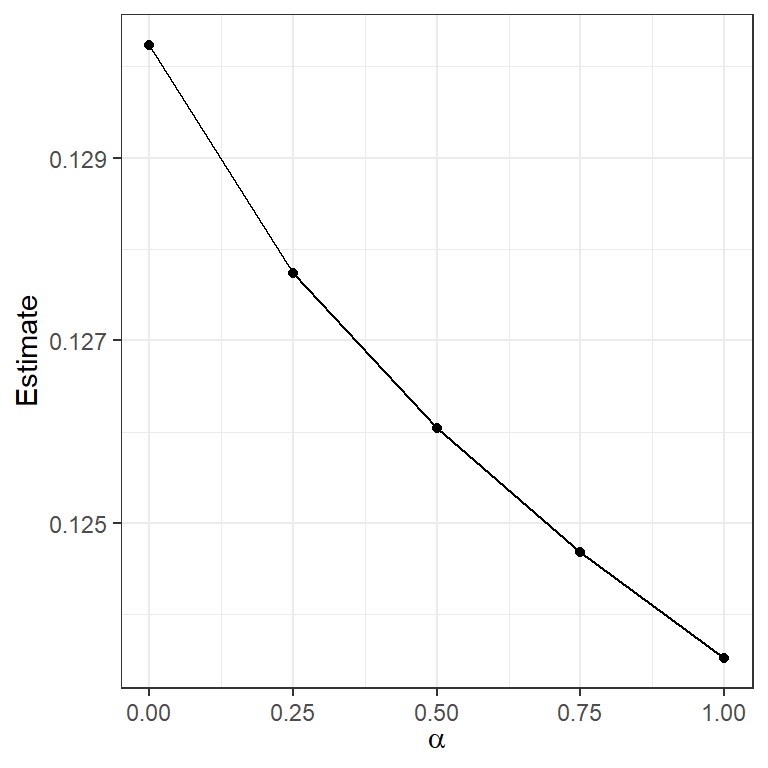}
        \caption{$\beta_{\text{BMS}}$}
    \end{subfigure}
    
    \caption{Evolution of the estimated coefficients as a function of the adjustment parameter $a$.}
    \label{fig:EvolBeta}
\end{figure}

For our numerical illustration, we restrict attention to the Exposure-Weighted specification, although the other approaches could also be applied. The panels in Figure~\ref{fig:EvolBeta} display the evolution of the components of the estimator $\boldsymbol{\widehat{\beta}}$ as a function of the adjustment parameter~$a$, with the intercept excluded from the figure for readability. This representation facilitates the visualization of how the mid-term cancellation penalty affects the relativities associated with the covariates. In particular, by examining the estimated values of $\beta_{\text{BMS}}$ across different values of $a$, one can observe that stronger penalties for early cancellations tend to reduce the influence of the claims experience. It can also be noted that the evolution of the estimators is not always monotonic, as illustrated, for instance, by the behavior of the estimate of $\beta_1$.

Finally, Table~\ref{fig:scorforalpha} presents the predictive performance on the test dataset, based on the deviance and the area between the concentration and Lorenz curves for each value of $a$. As can be observed, these scores tend to improve as $a$ increases. Therefore, $a = 100\%$ corresponds to the best predictive performance. It is also worth noting that $a = 100\%$ is precisely the value associated with the optimal penalty function.

\begin{table}[ht]
\centering
\begin{tabular}{l|ccccc}
\hline
&\multicolumn{5}{c}{Smoothing penalty factor $a$}\\
Score & 0\% & 25\% & 50\% & 75\% & 100\% \\ 
\hline
Area & 0.1237 & 0.0934 & 0.0678 & 0.0470 & 0.0306 \\ 
Deviance & 114.77 & 111.76 & 110.09 & 108.96 & 108.12 \\ 
\hline
\end{tabular}
\caption{Predictive scores for different values of the smoothing penalty factor $a$}
\label{fig:scorforalpha}
\end{table}

\subsection{Loss Allocation through Mid-Term Cancellation Penalties} 
\label{sectAlloc}

One important reason why policyholders may cancel their insurance coverage mid-term is the occurrence of a major claim—for example, when the insured vehicle becomes unusable and must be replaced. Within a pricing structure that penalizes mid-term cancellations, the insurer may design penalties that partially reflect the occurrence of such severe claims. In doing so, the insurer indirectly identifies large losses through the cancellation surcharge, thereby recovering part of the expected cost associated with the claim.

Let $\pi^{\mathrm{Flex}}$ denote the annual premium corresponding to $\gamma(1)$. Under the proposed model, the premium charged to a policyholder with exposure $t$ is expressed as $\gamma(t)\,\pi^{\mathrm{Flex}}$, where $\gamma(t)$ is the adjustment factor associated with mid-term cancellation behavior. This quantity naturally decomposes into two components:
\begin{enumerate}
    \item the \textit{pro rata} premium corresponding to the effective period of coverage, $t\pi^{\mathrm{Flex}}$;
    \item the additional amount associated with the cancellation \textit{penalty}, $\rho(t)$.
\end{enumerate}

As introduced in Section~\ref{sectPenalty}, this decomposition can be written as
\[
    \gamma(t)\,\pi^{\mathrm{Flex}} \;=\; t\,\pi^{\mathrm{Flex}} \;+\; \rho(t).
\]

Empirically, approximately 11\% of the total premium collected in the portfolio is attributable to this penalty component (10.89\% in the training set and 10.80\% in the test set). Moreover, as illustrated by the curves in Figure~\ref{mucomparison}, the proposed ratemaking structure naturally imposes a higher surcharge on policyholders who cancel their contracts mid-term than the traditional approach. More precisely, conditional on the $XO$ group (i.e., contracts canceled mid-term), the proposed model recovers approximately 15\% more premium from these policyholders compared with the traditional specification (15.28\% in the training set and 15.01\% in the test set).

\begin{figure}[H]
	\begin{center}
		\includegraphics[scale=0.54]{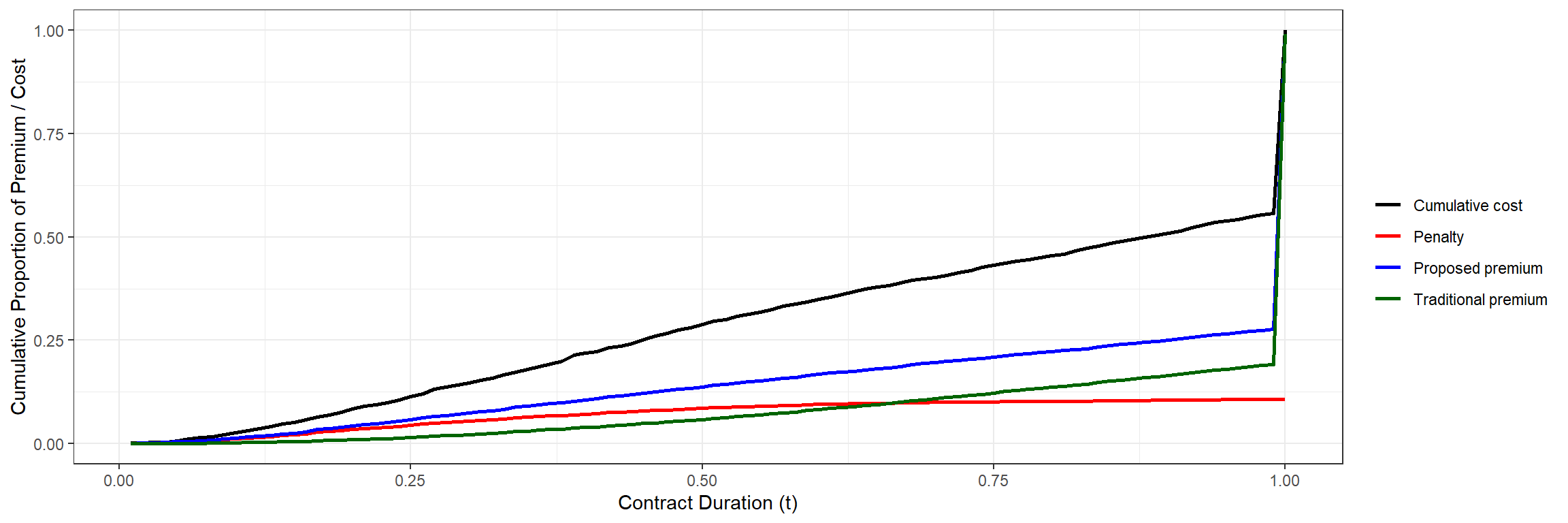}
	\caption{Cumulative proportion of premiums and costs by exposure level}
		\label{Allocation}
	\end{center}
\end{figure}

Finally, using the testing dataset, Figure~\ref{Allocation} displays the cumulative accumulation of premiums as a function of risk exposure $t$. The blue curve represents the proposed premium $\gamma(t)\,\pi$; the green curve corresponds to the traditional model; and the red curve shows the cumulative penalty $\rho(t)$, which is included in the blue curve. Comparison with the total claim costs (black curve) indicates that discrepancies in model fit remain---very likely caused by the pricing constraints C1 and C2 introduced in Section~\ref{sectPenalty}. This suggests that it may be beneficial for insurers to develop refined approaches capable of capturing a larger share of premium from policyholders who cancel their contracts mid-term.

\subsubsection{Competitive Market} 

A final perspective that deserves closer examination concerns the potential competitive advantage that such a model could offer to an insurer implementing it. The annual premium for full coverage is typically the main factor considered by policyholders when choosing an insurer, whereas the structure of mid-term cancellation penalties is unlikely to play a significant role in their decision. Consequently, increasing the mid-term cancellation penalties effectively leads to a reduction in the annual premium for full-year coverage.

\begin{figure}[H]
	\begin{center}
		\includegraphics[scale=0.54]{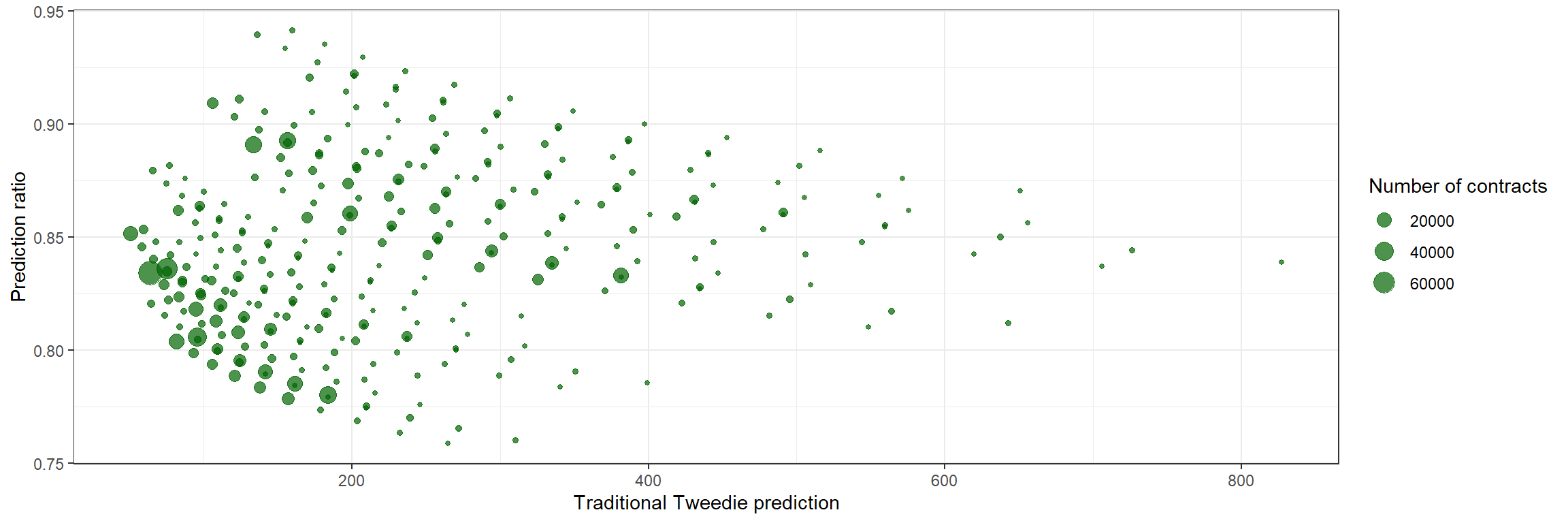}
	\caption{Annual premium ratio between the flexible and the traditional models}
		\label{ratio}
	\end{center}
\end{figure}

To further illustrate this phenomenon, Figure \ref{ratio} displays the ratio between the full-year premiums predicted by the proposed full-penalty approach ($a=1$) and those produced by the traditional model, evaluated across all policyholder profiles generated from the covariates used in this study. The size of the green dots reflects the number of contracts observed in the test dataset for each profile, while the red dots (of which there are very few) correspond to unobserved profiles for which the model nevertheless produces a premium estimate.  The figure clearly shows that the proposed approach yields substantially lower annual premiums, with ratios consistently below one. In some cases, the reduction is quite substantial: for certain profiles that typically generated annual premiums around \$300, the new pricing structure offers discounts of up to 25\%.

\subsection{Tweedie with Group-Specific Spline}

\begin{figure}[H]
	\begin{center}
		\includegraphics[scale=0.54]{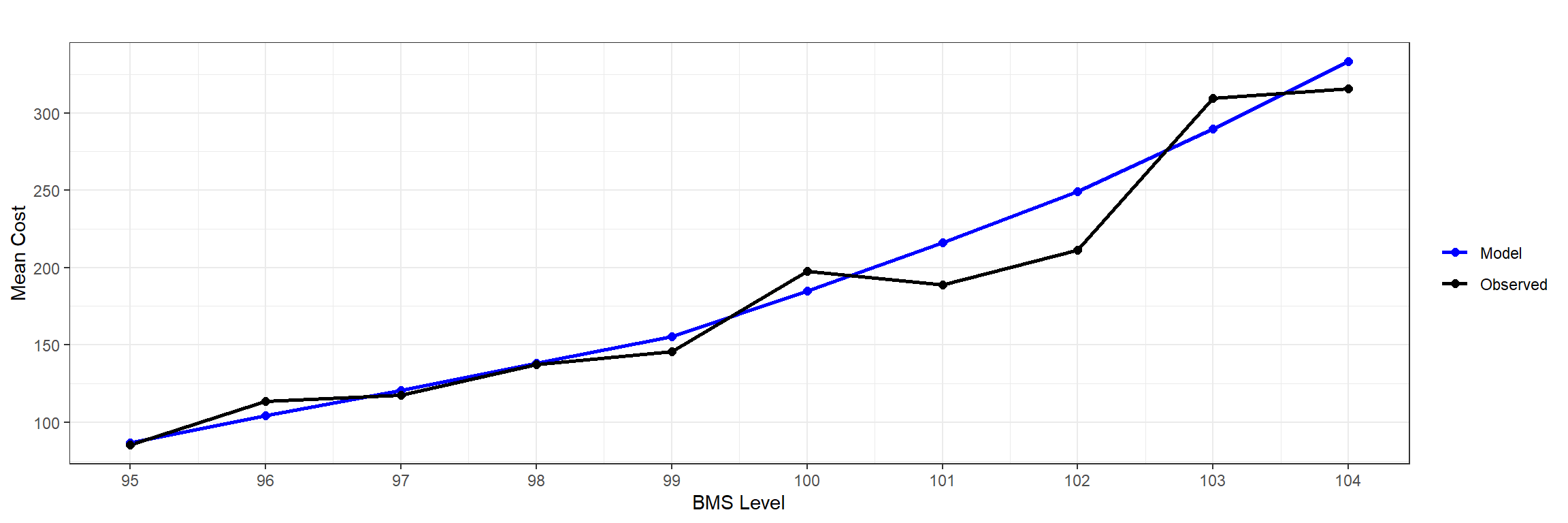}
	\caption{Comparison between observed average costs and predicted values across BMS levels}
		\label{BMS1}
	\end{center}
\end{figure}

Although the new ratemaking approaches that explicitly account for mid-term cancellations have produced promising results, there remains room for further investigation. In Section \ref{SectDataBMS}, we highlighted one policyholder characteristic that appears particularly relevant for understanding both loss costs and mid-term cancellations: the BMS level. As an initial step, Figure \ref{BMS1} presents a comparison between the predicted amounts from the proposed pricing model and the observed costs across BMS levels. Although the model fit appears reasonably good, it is well known that policyholders with higher BMS levels—corresponding to riskier profiles—tend to file more claims and are therefore more likely to cancel their policies mid-term following a loss event. This observation suggests that introducing group-specific spline functions based on the BMS level could represent a natural extension of the proposed framework. Such an approach would allow the penalty function to vary with policyholder risk characteristics, making it possible to allocate a slightly larger share of premium to higher-risk individuals who are statistically more prone to mid-term cancellations, thereby improving both fairness and profitability, as previously illustrated in Section \ref{sectAlloc}.

Our objective is therefore to develop a more general function $\gamma(t)$ that explicitly incorporates policyholder characteristics—using the BMS level as a first step in our application. By allowing the mid-term cancellation penalty to vary with individual attributes, the model can better capture heterogeneity in policyholder behavior, enhance predictive accuracy, and provide a more refined understanding of the relationship between exposure and observed outcomes.

\subsubsection{Grouping by BMS Level}

To generalize the smoothing model of the risk exposure according to the policyholder’s risk profile defined by the BMS level, we aim to partition policyholders into groups based on their BMS level. More formally, let $k \in \{1,2\}$ denote the BMS group associated with contract, and $t$ its exposure. We then consider a Tweedie-type model with a logarithmic link, in which the penalty function depends nonlinearly on exposure and varies across BMS groups. Specifically, the mean parameter is specified as:

\begin{equation}
\mu
= \exp\!\left\{ \mathbf{x}^\top \bm{\beta} 
+ f_{k}(t) \right\}, \quad k \in \{1,2\},
\label{eq:mu_bmsgroup}
\end{equation}

where $\mathbf{x}$ includes the six explanatory covariates used previously, $\bm{\beta}$ is the corresponding parameter vector, and $f_{1}(\cdot)$ and $f_{2}(\cdot)$ are two distinct smooth functions (penalized splines) of exposure $t$, estimated separately for groups~1 and~2.

In other words, instead of assuming a single function $\gamma_{\text{}}(t)$ common to the entire portfolio, we allow the shape of the relationship between exposure and expected cost to vary across risk profiles as defined by the BMS level. This specification enables the model to capture differentiated penalty structures for low- and high-risk policyholders, while maintaining the Tweedie generalized linear modeling framework.

\begin{figure}[H]
	\begin{center}
		\includegraphics[scale=0.54]{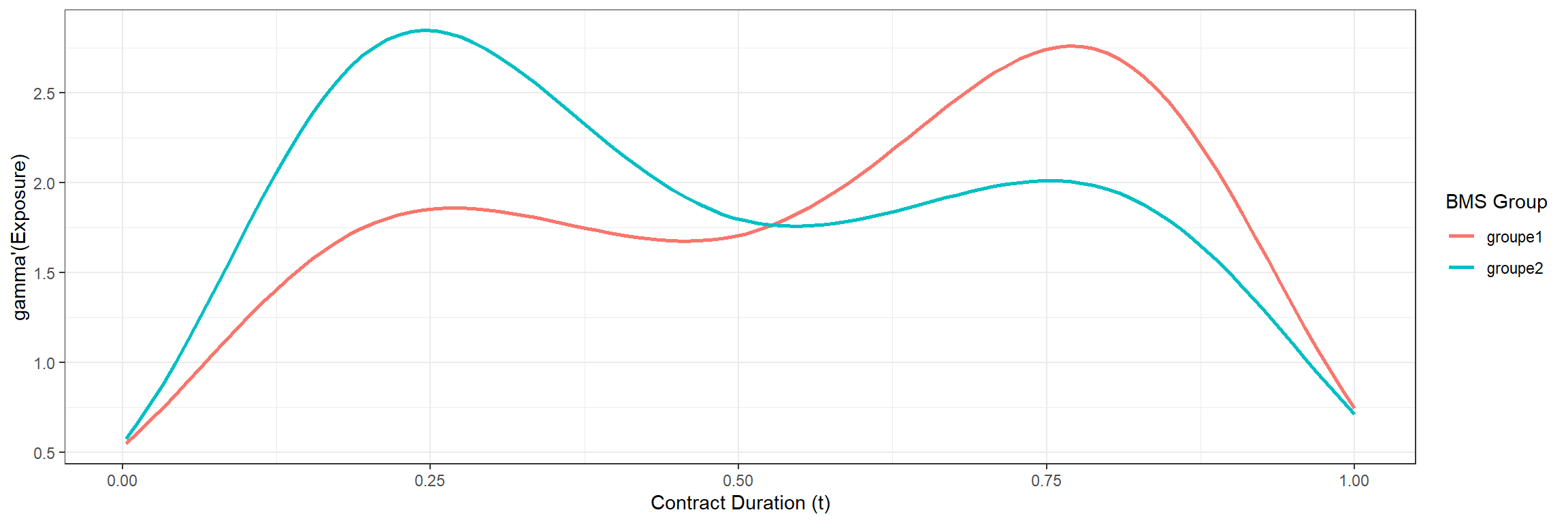}
	\caption{Estimated smooth functions $\gamma_{\text{con}}(d)$ by BMS group}
		\label{BMSGroup}
	\end{center}
\end{figure}

To identify the most appropriate grouping, we proceeded iteratively by dividing policyholders into two subsets according to their BMS level. Given the ordinal nature of the BMS level—where lower levels (e.g., 95) correspond to safer drivers and higher levels (e.g., 104) to riskier ones—we searched for the optimal cut point that maximizes the difference between the two estimated smooth functions. By testing all eight possible divisions, the best segmentation, illustrated in Figure~\ref{BMSGroup}, was obtained using the following BMS groups:

\begin{itemize}
    \item \textbf{Group 1}: levels $95 \leq \text{BMS} \leq 99$,
    \item \textbf{Group 2}: levels $100 \leq \text{BMS} \leq 104$.
\end{itemize}

\begin{figure}[H]
	\begin{center}
		\includegraphics[scale=0.54]{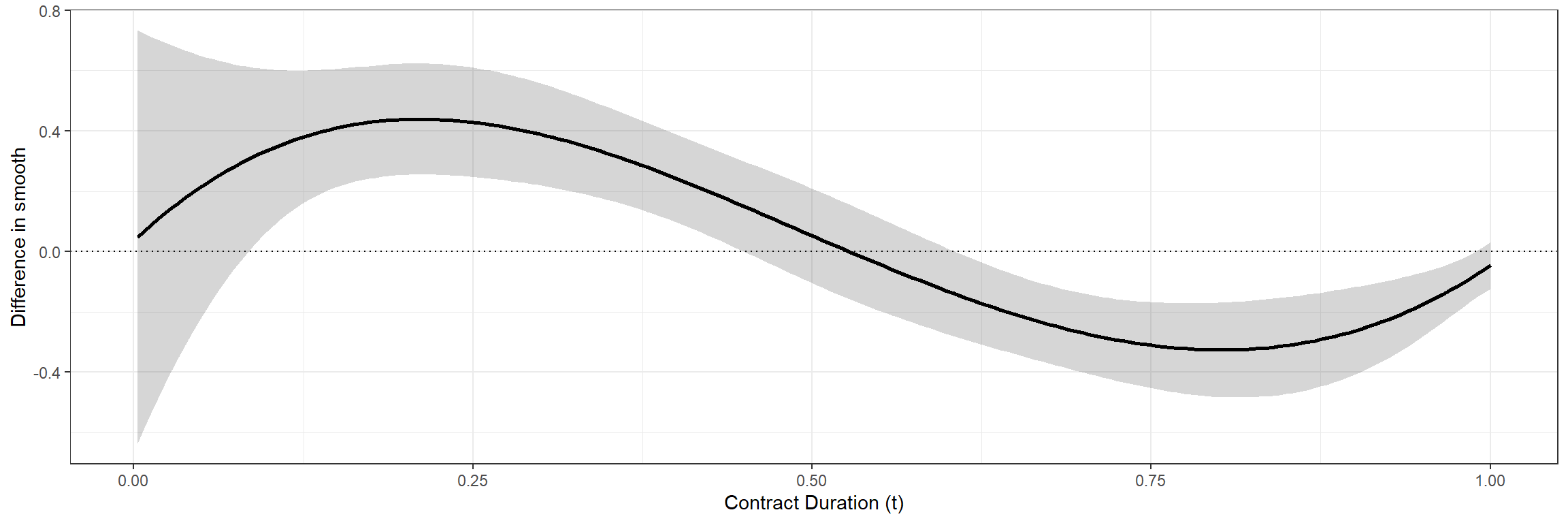}
	\caption{Difference between smooth functions for BMS groups}
		\label{BMSGroup2}
	\end{center}
\end{figure}

The difference between the estimated smooth functions for the two groups, together with the corresponding standard errors, is shown in Figure~\ref{BMSGroup2}. It is interesting to observe that the smooth function for Group~1 lies below that of Group~2 around mid-year, while it becomes higher thereafter. Except for the very beginning of the policy period and around mid-year, the smooth functions are statistically different, confirming that the effect of exposure varies significantly across BMS levels. Further attempts to subdivide either group did not yield additional statistically significant differences.

It is particularly interesting to note how the division between the two groups of policyholders emerged. Group~1 corresponds to policyholders who have maintained at least one claim-free year of insurance, while Group~2 includes both new policyholders and those who have filed a claim in recent years. In other words, this implies that insurers would benefit from imposing stronger mid-term cancellation penalties (as illustrated by the blue spline in Figure~\ref{BMSGroup}) for newly insured policyholders. Given that new policyholders typically represent a higher administrative cost segment, this form of penalty is also operationally relevant. For future work, it would be worthwhile to generalize the proposed framework by allowing the mid-term cancellation penalty to depend on additional covariates, thereby capturing more complex interactions between individual characteristics and cancellation behavior.

The next step in the practical application of an approach with penalty curves differentiated by policyholder profile—in this case, by BMS level—should involve applying the ratemaking constraints C1 and C2, as introduced in Section \ref{sectPenalty} and implemented in our initial pricing model in Section \ref{SectPenRate}, illustrated in Figure \ref{penalty1}. As shown in Figure \ref{BMSGroup}, enforcing a consistent penalty structure for policyholders in Group 2 would result in a very steep curve: the full annual premium would be charged after only one quarter of coverage. Although this may appear extreme, it is not entirely out of line with many one-year subscription-based products. In several industries—such as fitness memberships, telecommunications services, or annual software licenses—front-loaded fees or non-refundable upfront costs imply that a large fraction of the annual price is effectively charged within the first months of the contract. 

For policyholders in Group 1, the penalty structure would be more similar to the one developed for the initial model. A smoothing of the two penalty curves using a factor $a$, as proposed in Section \ref{SectSmoothPen}, could also be considered to achieve a more balanced adjustment between the groups.

\section{Conclusion} \label{Conclu}

This paper has proposed a new family of Tweedie-based ratemaking models that explicitly account for mid-term policy cancellations, addressing an aspect of insurance pricing that is often overlooked in traditional frameworks. By introducing a flexible penalty function $\gamma(t)$ and alternative weighting schemes, the proposed approach provides a more realistic representation of the earned premium when coverage is interrupted before policy maturity. Through a combination of theoretical developments and empirical analysis, we have shown that the specification of both the mean and weight functions has a significant impact on parameter estimation and model stability.

Empirical results based on real automobile insurance data confirm that policyholders who cancel their coverage mid-term exhibit higher claim frequencies and severities, reinforcing the need to incorporate cancellation behavior into ratemaking models. The proposed constrained Tweedie models, together with the adjustment penalty parameter $a$, demonstrate that it is possible to reconcile statistical performance with practical interpretability, offering a flexible mechanism to penalize early cancellations while preserving fairness toward policyholders who maintain full-year coverage.

Beyond the models developed in this paper, several extensions could be explored to further enhance the flexibility and interpretability of the proposed framework. In particular, the penalty function $\gamma(t)$—currently estimated as a single smooth function of exposure—could itself be modeled as varying across policyholder characteristics. A first attempt in this direction was made by allowing distinct spline functions for different BMS levels. This specification implicitly assumes that mid-term cancellation penalties differ according to the policyholder’s risk profile, which is consistent with actuarial intuition: safer drivers should face less stringent penalties than higher-risk individuals.

Preliminary results suggest that this extension is both statistically and conceptually promising. By conditioning the shape of $\gamma(t)$ on covariates such as the BMS level or driving experience, it becomes possible to capture heterogeneous behavioral responses to mid-term cancellations and to refine the alignment between premium structure, risk exposure, and observed claim patterns. Future research will aim to formalize this multi-dimensional modeling of the penalty spline and to assess its implications for risk classification, portfolio segmentation, and insurer competitiveness.

\section*{Acknowledgements}

The two first authors are grateful to the Natural Sciences and Engineering Research Council of Canada (NSERC) for the financial support provided by the NSERC Alliance grant supporting the \textit{Chaire Co-operators en analyse des risques actuariels} (Grant No. ALLRP 597285-2024). Jean-Philippe Boucher further acknowledge the NSERC for the financial support through the Discovery Grant program (RGPIN-2025-03995).

\bibliography{bibtex}

@article{delong2021making,
  title={Making Tweedie’s compound Poisson model more accessible},
  author={Delong, {\L}ukasz and Lindholm, Mathias and W{\"u}thrich, Mario V},
  journal={European Actuarial Journal},
  volume={11},
  number={1},
  pages={185--226},
  year={2021},
  publisher={Springer}
}

@book{frees2014predictive,
  title={Predictive modeling applications in actuarial science},
  author={Frees, Edward W and Derrig, Richard A and Meyers, Glenn},
  volume={1},
  year={2014},
  publisher={Cambridge University Press}
}

@book{denuit2007actuarial,
  title={Actuarial modelling of claim counts: Risk classification, credibility and bonus-malus systems},
  author={Denuit, Michel and Mar{\'e}chal, Xavier and Pitrebois, Sandra and Walhin, Jean-Fran{\c{c}}ois},
  year={2007},
  publisher={Wiley, West Sussex}
}

@article{raissaboucher2024, 
title={Bonus-Malus Scale premiums for Tweedie’s compound Poisson models}, 
DOI={10.1017/S1748499524000113}, 
journal={Annals of Actuarial Science}, 
author={Boucher, Jean-Philippe and Coulibaly, Raïssa}, 
pages={1--25},
year={2024}
}

@article{raissaboucher2025,
  author  = {Boucher, Jean-Philippe and Coulibaly, Raissa},
  title   = {Comparison of offset and ratio weighted regressions in Tweedie models with application to mid-term cancellations},
  journal = {European Actuarial Journal},
  year    = {2026},
  doi     = {10.1007/s13385-026-00452-z}
}

@article{nelder1972generalized,
  title={Generalized linear models},
  author={Nelder, John Ashworth and Wedderburn, Robert WM},
  journal={Journal of the Royal Statistical Society Series A: Statistics in Society},
  volume={135},
  number={3},
  pages={370--384},
  year={1972},
  publisher={Oxford University Press}
}

@inproceedings{tweedie1984index,
  title={An index which distinguishes between some important exponential families},
  author={Tweedie, Maurice CK and others},
  booktitle={Statistics: Applications and new directions: Proc. Indian statistical institute golden Jubilee International conference},
  volume={579},
  pages={579--604},
  year={1984}
}

@book{jorgensen1997theory,
  title={The theory of dispersion models},
  author={J{\o}rgensen, Bent},
  year={1997},
  publisher={CRC Press}
}

@book{casella2024statistical,
  title={Statistical inference},
  author={Casella, George and Berger, Roger},
  year={2024},
  publisher={CRC Press}
}

@article{denuit2024testing,
  title={Testing for auto-calibration with Lorenz and Concentration curves},
  author={Denuit, Michel and Huyghe, Julie and Trufin, Julien and Verdebout, Thomas},
  journal={Insurance: Mathematics and Economics},
  volume={117},
  pages={130--139},
  year={2024},
  publisher={Elsevier}
}

@article{denuit2024convex,
  title={Convex and Lorenz orders under balance correction in nonlife insurance pricing: Review and new developments},
  author={Denuit, Michel and Trufin, Julien},
  journal={Insurance: Mathematics and Economics},
  year={2024},
  publisher={Elsevier}
}

@article{white1982maximum,
  title={Maximum likelihood estimation of misspecified models},
  author={White, Halbert},
  journal={Econometrica: Journal of the econometric society},
  pages={1--25},
  year={1982},
  publisher={JSTOR}
}

@book{wood2017,
  title={Generalized additive models: an introduction with R},
  author={Wood, Simon N},
  year={2017},
  publisher={Chapman and Hall/CRC}
}

@article{denuit2019model,
  title={Model selection based on Lorenz and concentration curves, Gini indices and convex order},
  author={Denuit, Michel and Sznajder, Dominik and Trufin, Julien},
  journal={Insurance: Mathematics and Economics},
  volume={89},
  pages={128--139},
  year={2019},
  publisher={Elsevier}
}

@article{pechon2019multivariate,
  title={Multivariate modelling of multiple guarantees in motor insurance of a household},
  author={Pechon, Florian and Denuit, Michel and Trufin, Julien},
  journal={European Actuarial Journal},
  volume={9},
  pages={575--602},
  year={2019},
  publisher={Springer}
}

@book{lemaire,
  title={Bonus-malus systems in automobile insurance},
  author={Lemaire, Jean},
 year={2012},
volume =  {19},
 publisher={Springer science}
}

@article{scoringrules,
  title={Strictly proper scoring rules, prediction, and estimation},
  author={Gneiting, Tilmann and Raftery, Adrian E},
  journal={Journal of the American statistical Association},
  volume={102},
  number={477},
  pages={359--378},
  year={2007},
  publisher={Taylor \& Francis}
}

@article{turcotte2022gamlss,
  title={GAMLSS for Longitudinal Multivariate Claim Count Models},
  author={Turcotte, Roxane and Boucher, Jean P},
  journal={North American Actuarial Journal},
  pages={1--24},
  year={2023},
  publisher={Taylor \& Francis}
}

@article{denuit2025comparison,
  title   = {Comparison of predictors’ performance in insurance pricing: testing for Bregman dominance based on Murphy diagrams},
  author  = {Denuit, Michel and Trufin, Julien and Verdebout, Thomas},
  journal = {European Actuarial Journal},
  volume  = {15},
  number  = {2},
  pages   = {493--504},
  year    = {2025},
  publisher = {Springer}
}

\newpage
\appendix
\renewcommand{\thesection}{Appendix \Roman{section}: }
\section{Descriptive statistics of covariates}
\label{SectAppendix1}

\begin{figure}[H]
	\begin{center}
		\includegraphics[scale=0.65]{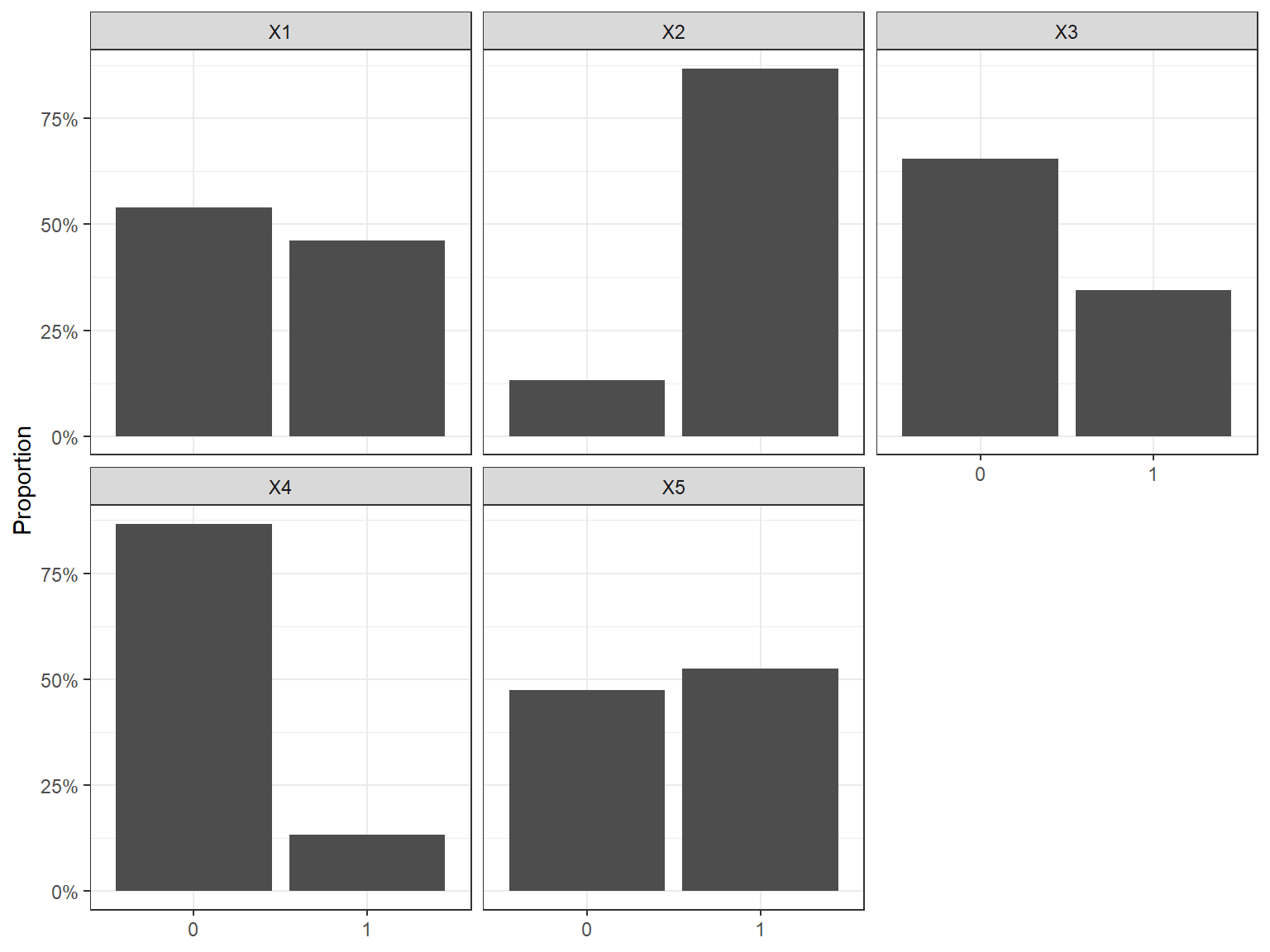}
	\caption{Descriptive statistics of all 5 covariates from the database}
		\label{cov.stat}
	\end{center}
\end{figure}

Figure \ref{cov.stat} provides a brief overview of the five covariates included in the Tweedie models. Each variable has been standardized and coded into two categories (0 and 1) for simplicity of interpretation. The table reports the proportion of observations in each category across the training dataset.

\newpage
\section{Weight iterations}
\label{SectAppendix2}

\begin{figure}[H]
	\begin{center}
		\includegraphics[scale=0.50]{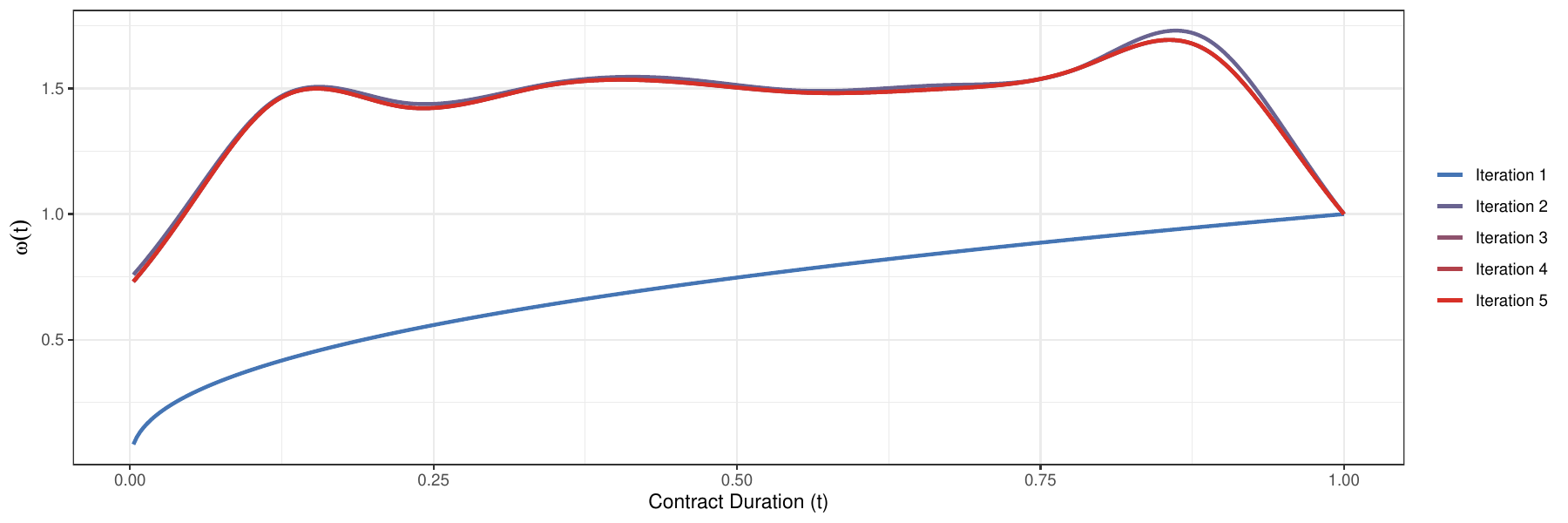}
	\caption{Evolution of the weight function $\omega (t)$ over the iterative estimation process}
		\label{Appendix}
	\end{center}
\end{figure}

Figure \ref{Appendix} illustrates the evolution of the weight function $\omega_i$ across the successive iterations of the estimation algorithm described in Section \ref{Tweediedistribution}. The figure provides a graphical representation of how the spline-based approximation of $\gamma(d_i)$ stabilizes as the iterative procedure updates both the mean and weight components. Each curve corresponds to a given iteration $k$, starting from the initial Exposure Weighted Model weighting scheme. A small horizontal offset was applied to each curve to improve visual distinction. The rapid convergence observed after only a few iterations confirms the numerical stability and internal consistency of the Gamma Weighted estimation process.

\end{document}